\documentclass[aps,prd,twocolumn,superscriptaddress,nofootinbib,showpacs,showkeys]{revtex4-1}
\usepackage{graphicx}  
\usepackage{dcolumn}   
\usepackage{bm}        
\usepackage{amssymb}   
\usepackage{amsmath}   
\usepackage{lineno}
\usepackage{draftcopy}
\usepackage{relsize} 
\usepackage{xfrac} 
\usepackage{slashed}   

\usepackage{color}
\definecolor{dgreen}{cmyk}{1.,0.,1.,0.2}        
\definecolor{orange}{cmyk}{0.,0.353,1.,0.}    

\usepackage[bookmarks]{hyperref}

\newcommand{\di}{{\rm d}}

\newcommand{\be}{\begin{equation}}
\newcommand{\ee}{\end{equation}}                                                                               
\newcommand{\bea}{\begin{eqnarray}}
\newcommand{\eea}{\end{eqnarray}}

\begin{document}
\title{"Splitting" magnetic catalysis effect prevents vacuum superconductivity in strong magnetic fields}
\author{Gaoqing Cao}
\affiliation{School of Physics and Astronomy, Sun Yat-Sen University, Guangzhou 510275, China.}
\date{\today}

\begin{abstract}
By comparing the two- and three-flavor Nambu--Jona-Lasinio (NJL) models, we demonstrate that the naively expected vacuum superconductivity (VSC) in constant magnetic field ${\bf B}=B\hat z$ is disfavored due to the splitting magnetic catalysis effect (MCE) to chiral condensates with different quark flavors. Based on the simple two-flavor NJL model, we illuminate, in the lowest Landau level approximation, the similar origins of $\pi^0$ and $\bar{\rho}^+_1$ (${\rho}^+$ meson with spin $S_z=1$) mass reductions with smaller $B$ and their different features at larger $B$. With the full Landau levels, the two-flavor NJL model is found to be invalid to study the magnetic field effect to $\bar{\rho}^+_1$ meson with physical vacuum mass $775~{\rm MeV}$. Then, restricted to $\rho$ meson mass below two-quark threshold in vacuum, that is $m_\rho^v<2m_q^v$, it is found that $\pi^0$ mass decreases and then increases with $B$ slowly, and $\bar{\rho}^+_1$ mass vanishing point is delayed to larger $B$ compared to the point particle result. In the more realistic three-flavor NJL model, all the quark masses split in strong magnetic field as a combinatorial result of their different current masses and electric charges. By choosing a vacuum mass closer to the physical one, $\bar{\rho}^+_1$ meson mass is found to be consistent with the LQCD results semi-quantitatively in smaller $B$ region but increase in larger $B$ region. These features are mainly outcomes of the interplay between the $S_z-B$ coupling effect and splitting MCE to the composite $u$ and $d$ quarks, which definitely disfavors VSC when the latter dominates. Furthermore, mesonic flavor mixing is modified by $B$ among the neutral pseudoscalars: $\pi^0,\eta_0$ and $\eta_8$, which is very important to suppress the mass enhancement of the effective mass eigenstates at large $B$.
\end{abstract}

\maketitle

\section{Introduction}
The properties of quantum chromodynamics (QCD) system in external electromagnetic (EM) field are very interesting and significant in both theoretical and experimental aspects. Theoretically, many novel notions and possibly new physics emerge from such system, such as, macroscopic chiral anomaly effects~\cite{Kharzeev:2010gd,Son:2004tq,Metlitski:2005pr,Huang:2013iia,Hattori:2016njk}, inverse magnetic catalysis effect (IMCE)~\cite{Bali:2011qj,Bali:2012zg,Bruckmann:2013oba,Fukushima:2012kc,Kojo:2012js,Chao:2013qpa,Cao:2014uva,Ferrer:2014qka,Mao:2016fha} and vacuum superconductivity~\cite{Chernodub:2010qx,Chernodub:2011mc,Braguta:2011hq,Liu:2014uwa,Hidaka:2012mz,Bali:2017ian}. Experimentally, the strongest EM field in our recent Universe can be produced in relativistic peripheral heavy ion collisions~\cite{Skokov:2009qp,Deng:2012pc,Deng:2014uja,Bloczynski:2012en,Guo:2019joy} and chiral magnetic effect~\cite{Liao:2014ava,Kharzeev:2015znc,Huang:2015oca} is now under restrict and massive investigations in BES II of STAR experiments~\cite{Adamczyk:2014mzf,Zhao:2017nfq,Magdy:2017yje}. Focusing on the theoretical part, the external EM field actually contributes an extra dimension, besides the usual temperature and chemical potential effect, to explore the properties of QCD especially the phase diagrams. First of all, it is important to emphasize that the first-principle lattice QCD (LQCD) simulations~\cite{Bali:2011qj,Bali:2012zg} greatly support the MCE in vacuum, that is, the chiral condensations (or likely quark masses) increase with magnetic field~\cite{Miransky:2015ava}. Then in the phase respect, inhomogeneous chiral symmetry breaking ($\chi$SB) phases might be favored for finite density system in the presence of magnetic field~\cite{Frolov:2010wn,Cao:2016fby}, vacuum superconductivity is assumed to happen at large enough magnetic field~\cite{Chernodub:2010qx,Chernodub:2011mc,Braguta:2011hq,Liu:2014uwa}, and neutral pseudoscalar superfluidity can be found in parallel EM field~\cite{Cao:2015cka,Wang:2017pje,Wang:2018gmj}. 

Based on the ordinary $\chi$SB phase, the meson masses were further studied in magnetic field, either neutral or charged. Most frequently worked out in two-flavor NJL model: neutral pion mass is found to decrease and then increase, and charged pion mass to increase monotonously with magnetic field~\cite{Wang:2017vtn,Mao:2018dqe,Liu:2018zag,Avancini:2016fgq,Coppola:2018vkw}, which then both disfavor pion superfluidity in consistent with the restriction from the Gell-Mann--Oakes--Renner relation~\cite{Shushpanov:1997sf,Agasian:2001ym}; while the lightest charged vector rho meson mass decreases monotonously with $B$ to zero, which favors vacuum superconductivity~\cite{Chernodub:2010qx,Chernodub:2011mc,Braguta:2011hq,Liu:2014uwa}. There are both quenched~\cite{Bali:2017ian} and unquenched LQCD simulations~\cite{Hengtong2019} on the pion masses in the market recently: while the charged pion mass trivially increases with $B$, the neutral pion mass decreases to an almost convergent value, around half of the vacuum mass. It seems a {\it puzzle} why neutral pion mass will converge to such a specific value. The initial philosophy of VSC simply follows the expectation from point vector particle, whose effective mass $m_V^B=\sqrt{{m_V^v}^2-|eB|}$ vanishes at $|eB|={m_V^v}^2$. The most prominent example of the latter is the electroweak gauge boson $W^\pm$ condensation in early Universe~\cite{Ambjorn:1988tm,Ambjorn:1988gb}, which was later shown to exhibit superconductivity~\cite{Chernodub:2010qx,Chernodub:2012fi}. However, the proposal of VSC encounters strong objections from the community due to the violation of Vafa-Witten (VW) theorem and was denied by LQCD simulations~\cite{Hidaka:2012mz,Bali:2017ian}. It composes the main motivation of this work to find out which ingredient or underlying physics is missing in the two-flavor NJL model in order to account for the contradiction with LQCD results. To avoid confusion, we just focus on the possible continuum phase transition to homogeneous VSC, for which the Ginzburg-Landau expansion and thus the criteria with zero-mass point as the transition point are valid~\cite{Cao:2015xja}.

The paper is organized as the following: In Sec.~\ref{2flavor}, we develop the whole formalism for the explorations of $\pi^0$ and $\rho$ meson masses under strong magnetic field within extended two-flavor NJL model. For the purpose of an intuitive understanding, we show the similarity in their origins between the mass reductions of $\pi^0$ and $\bar{\rho}^+_1$ at smaller $B$ by adopting lowest Landau level (LLL) approximation in Sec.~\ref{LLL}. Then, the full Landau level (FLL) expressions are given explicitly in Sec.~\ref{FLL} together with the associated numerical calculations. Based on this two-flavor model, the important discussions on the equality between the proper-time and Landau-level presentations and the invalidity of NJL model study of $\rho$ meson with physical mass in magnetic field are reserved in Appendix.~\ref{equality} and Appendix.~\ref{invalidity}, respectively. In Sec.~\ref{3flavor}, we revisit the meson modes in the more realistic three-flavor NJL model and present the FLL numerical results.

\section{Meson spectra within two-flavor Nambu--Jona-Lasinio model}\label{2flavor}

In order to study the properties of vector mesons, the original Lagrangian density of the two-flavor Nambu--Joan-Lasinio (NJL) model~\cite{Klevansky:1992qe} can be extended by including vector channels and keeping approximate chiral symmetry to~\cite{Chernodub:2011mc,Liu:2014uwa}
\begin{eqnarray}\label{njl2}
{\cal L}&=&\bar \psi\left(i\slashed D-m_0\right)\psi+G_S\left[\left(\bar\psi\psi\right)^2+\left(\bar\psi i\gamma_5\boldsymbol\tau\psi\right)^2\right]\nonumber\\
&&-G_V\left[\left(\bar\psi\gamma^\mu\tau^a\psi\right)^2+\left(\bar\psi i\gamma^\mu\gamma_5\tau^a\psi\right)^2\right].
\end{eqnarray}
Here, $\psi=\left(u,d\right)^T$ is the two-flavor quark field, $m_0$ is the current quark mass, $\tau^a=(1,\boldsymbol\tau)$ with $\boldsymbol\tau$ pauli matrices in flavor space, and $G_S$ and $G_V$ are positive coupling constants for the scalar-pseudoscalar and vector-pseudovector channels, respectively. The covariant derivative $D_\mu=\partial_\mu+iq A_\mu$ is defined in flavor space with electric charge $q_{\rm u}=2e/3~(q_{\rm d}=-e/3)$ for $u~(d)$ quark and the vector potential $A_\mu=(0,0,-Bx_1,0)$ representing a constant magnetic field along $z$-axis through ${\bf B}=\nabla\times{\bf A}=B\hat z$. For the convenience of exploring the properties of the collective excitation modes or mesons, we introduce the following auxiliary boson fields:
\bea
&&\ \ \ \,\sigma=-2G_S\bar\psi\psi, \ \ \ \ \ \ \ \ \ \ \,{\boldsymbol \pi}=-2G_S\bar\psi i\gamma_5{\boldsymbol \tau}\psi,\\
&&V^{\mu a}=-2G_V\bar\psi\gamma^\mu\tau^a\psi, \ A^{\mu a}=-2G_V\bar\psi i\gamma^\mu\gamma_5\tau^a\psi.
\eea
Then, the Lagrangian density becomes~\cite{Chernodub:2011mc,Liu:2014uwa}
\begin{eqnarray}\label{HS}
\!\!\!{\cal L} &=& \bar\psi\left[i\tilde{\slashed {D}}-m_0-\sigma-i\gamma_5\left(\tau_3\pi^0+\tau_\pm\pi^\pm\right)\right]\psi-\nonumber\\
\!\!&&\!\!\!\!\!\!{\sigma^2\!\!+\!(\pi^0)^2\!\!+\!\pi^\mp\pi^\pm\over 4G_S}+{(\omega^\mu)^2\!\!+\!(\rho_{0}^{\mu})^2\!\!+\!\rho^{\mp}_{\mu}\rho^{\pm\mu}\!\!+\!(A^{a\mu})^2\over4G_V},\nonumber\\
&\tilde{D}_\mu&=\partial_\mu+i(q A_\mu\!-\!\omega_\mu\!-\!\tau_3\rho_{0\mu}\!-\!\tau_\pm\rho^{\pm\mu}\!-\!i\gamma_5\tau^aA_\mu^a),
\end{eqnarray}
where the physical fields are related to the auxiliary fields through: $\pi^0\!=\!\pi_3$, $\pi^\pm\!=\!{1\over\sqrt 2}\left(\pi_1\mp i\pi_2\right)$, $\rho_{0\mu}\!=\!\rho_{3\mu}$ and $\rho^{\pm}_{\mu}\!=\!{1\over\sqrt 2}\left(\rho_{1\mu}\mp i\rho_{2\mu}\right)$, and $\tau_\pm\!=\!{1\over\sqrt 2}\left(\tau_1\pm i\tau_2\right)$ are the raising and lowering operators in flavor space. If only the expectation value $\langle\sigma\rangle$ is nonzero as is the case in the vacuum without $B$, the thermodynamic potential is simply in the following form 
\begin{eqnarray}
\Omega={(m-m_0)^2\over 4G_S}+{i\over V_4}\sum_{\rm f=u,d}{\rm Tr}\ln G_{\rm f}^{-1},
\end{eqnarray}
where the dynamical mass $m=m_0+\langle\sigma\rangle$, $G_{\rm f}^{-1}=i\slashed D_{\rm f}-m$ is the inverse quark propagator at mean field level and the trace "${\rm Tr}$" should be taken over space-time coordinate, Dirac spinor, flavor and color spaces. 

Then, the gap equation is formally given by the minimum condition $\partial\Omega/\partial m=0$ as
\begin{eqnarray}\label{Fgap}
{m-m_0\over 2G_S}-{i\over V_4}\sum_{\rm f=u,d}{\rm Tr}~G_{\rm f}=0,
\end{eqnarray}
and the inverse meson propagators can be conveniently evaluated in random phase approximation (RPA) through~\cite{Klevansky:1992qe,Klimt:1989pm}:
\bea
D^{-1}_{SS}(y,x)&=&-{e^{-iq_S\int_{x}^yA\cdot\di x}\over2G_S}+{i\over V_4}{\rm Tr}~ {\cal G}\Gamma_{S^*} {\cal G}\Gamma_{S},\label{Sprp}\\
D^{-1}_{\bar{V}_{\mu}\bar{V}_{\nu}}(y,x)&=&{e^{-iq_V\int_{x}^yA\cdot\di x}g_{{\mu}{\nu}}\over2G_V}+{i\over V_4}{\rm Tr}~ {\cal G}\Gamma_{{\bar{V}_{\mu}}^*} {\cal G}\Gamma_{\bar{V}_{\nu}},\label{Vprp}
\eea
where ${\cal G}={\rm diag}(G_{\rm u},G_{\rm d})$ is the fermion propagator in flavor space, and $\Gamma_{S/S^*}$ and $\Gamma_{\bar{V}_{\mu}/\bar{V}_{\mu}^*}$ are the coupling vertices in scalar-pseudoscalar and vector-pseudovector channels, respectively. The explicit forms of the interested coupling vertices can be read from Eq.(\ref{HS}) as
\bea
&&\Gamma_{\sigma/\sigma^*}=-1,~\Gamma_{\pi^0/{\pi^0}^*}=-i\gamma^5\tau_3,~\Gamma_{\pi_\pm}=-i\gamma^5\tau_\pm,\nonumber\\
&&\Gamma_{\bar{\omega}_\mu/\bar{\omega}_\mu^*}=\bar{\gamma}_\mu^\pm,~\Gamma_{\bar{\rho}_{0\mu}/\bar{\rho}_{0\mu}^*}=\bar{\gamma}_\mu^\pm\tau_3,~\Gamma_{\bar{\rho}_{\pm\mu}}=\bar{\gamma}_\mu^\pm\tau_\pm,
\eea
where $\bar{\gamma}^{\pm}_\mu=(\gamma_0,{\gamma_1\pm i\gamma_2\over \sqrt{2}},{\gamma_1\mp i\gamma_2\over \sqrt{2}},\gamma_3)$ and the spin eigenstate $\bar{V}_{\mu}/\bar{V}_{\mu}^*=({V}_{0},{{V}_{1}\mp i{V}_{2}\over \sqrt 2},{{V}_{1}\pm i{V}_{2}\over \sqrt 2},V_3)$ with the spatial components $\bar{V}_{1}, \bar{V}_{2}$ and $\bar{V}_{3}$ corresponding to spin components $S_z=1,-1$ and $0$ along ${\bf B}$. $\bar{V}_{\mu}/\bar{V}_{\mu}^*$ are more convenient for the exploration of pole masses in magnetic field because $D^{-1}_{\bar{V}_{\mu}\bar{V}_{\nu}}$ vanishes at zero effective momentum if $\mu\neq\nu$. On thing should be pointed out: for nonlocal meson propagators, the Schwinger phases should be compensated for the charged mesons in order to keep gauge invariance of the theory in external EM field~\cite{Cao:2015xja}, see the Wilson lines in the first terms of Eq.\eqref{Sprp} and Eq.\eqref{Vprp} with the integral along a straight line. Then, their masses should be evaluated in energy-momentum space by taking out the gauge dependent Schwinger phases, that is, from
\bea
\!\!\!\!\!\!\!\!D^{-1}_{SS}(p)&\equiv&{1\over2G_S}+\Pi_{SS}(p)\nonumber\\
&=&\int\di^4x~e^{-ip\cdot (y-x)}e^{iq_S\int_{x}^yA\cdot\di x}D^{-1}_{SS}(y,x),\label{SprpP}\\
\!\!\!\!\!\!\!\!D^{-1}_{\bar{V}_{\mu}\bar{V}_{\mu}}(p)&\equiv&{1\over2G_V}+\Pi_{\bar{V}_{\mu}\bar{V}_{\mu}}(p)\nonumber\\
&=&\int\di^4x~e^{-ip\cdot (y-x)}e^{iq_V\int_{x}^yA\cdot\di x}D^{-1}_{\bar{V}_{\mu}\bar{V}_{\mu}}(y,x)\label{VprpP}
\eea
by requiring $D^{-1}(p_0,{\bf p}={\bf 0})=0$.
Thus obtained effective inverse meson propagators are equivalent to those directly evaluated with the effective quark propagators $S_{\rm f}({k})$, which will be defined immediately. 

\begin{widetext}
The basic quark propagators $G_{\rm f}(x,y)$ can be evaluated with Schwinger approach~\cite{Schwinger:1951nm} and we have
\begin{eqnarray}\label{qprp}
G_{\rm f}(x,y) &=& e^{-iq_{\rm f}\int_y^x A_{\rm f}^\mu dx_\mu}S_{\rm f}(x-y),\\
S_{\rm f}(x) &=& -i\int_0^\infty{ds\over 16(\pi s)^2}e^{-i\left[sm^2+{1\over 4s}\left(x_0^2-x_3^2-(x_1^2+x_2^2) B_{\rm f}^s \cot B_{\rm f}^s\right)\right]}
 B_{\rm f}^s\left[\cot B_{\rm f}^s+\gamma_1\gamma_2\right]\nonumber\\
 &&\left[m+{1\over 2s}\left(\slashed x_0-\slashed x_3-B_{\rm f}^s\left(\left(\slashed x_1+\slashed x_2\right)\cot B_{\rm f}^s-\slashed x_{21}+\slashed
   x_{12}\right)\right)\right]\nonumber
\end{eqnarray}
with $B_{\rm f}^s=q_{\rm f}Bs$, $\slashed x_\mu=\gamma_\mu x_\mu$, $\slashed x_{\mu\nu}=\gamma_\mu x_\nu$, and the integration in the exponential from $y$ to $x$ along a straight line. For later use, we shift to imaginary proper time $s\rightarrow-i\,s$ and transform the effective propagator $S_{\rm f}(x)$ to Euclidean energy-momentum space~\cite{Cao:2015xja}:
\begin{eqnarray}\label{SE}
S_{\rm f}({k}) = -i\int_0^\infty ds e^{-s\left(m^2+k_4^2+k_3^2+{\bf k}_\bot^2 {\tanh B_{\rm f}^s\over B_{\rm f}^s}\right)}\left[-\slashed k+m+i(\slashed k_{12}-\slashed k_{21})\tanh B_{\rm f}^s\right]\left(1+i\gamma_1\gamma_2\tanh B_{\rm f}^s\right)
\end{eqnarray}
with ${\bf k}_\bot=(k_1,k_2)$. Inserting the explicit quark propagators into Eq.\eqref{Fgap} and taking the vacuum regularization scheme with three-momentum cutoff, we finally have the finite gap equation~\cite{Cao:2015xja}
\begin{eqnarray}\label{gapm}
0&=& {m-m_0\over 2G}-{N_c m^2\over\pi^2}\left[\Lambda\sqrt{1+{\Lambda^2\over m^2}}-m\ln\left({\Lambda\over m}
+\sqrt{1+{\Lambda^2\over m^2}}\right)\right]-{N_cm\over 4\pi^2}\sum_{\rm f=u,d}\int_0^\infty{ds\over s^2} e^{-sm^2}\left({B_{\rm f}^s\over \tanh B_{\rm f}^s}-1\right).
\end{eqnarray}
\end{widetext}

\subsection{Intuition in lowest Landau level approximation}\label{LLL}

It was found in the previous explorations that both $\pi^0$ and $\bar{\rho}^+_1$ meson masses decrease with magnetic field in the weak field region~\cite{Chernodub:2011mc,Liu:2014uwa,Wang:2017vtn,Mao:2018dqe,Liu:2018zag,Avancini:2016fgq}, which may indicate neutral pion superfluidity (NPSF) and VSC, respectively, at sufficient strong magnetic field. In order to get an intuitive understanding of the situations encountered by $\pi^0$ and $\bar{\rho}^+_1$ mesons, we adopt the lowest Landau level (LLL) approximation with the effective quark propagators simply given by~\cite{Miransky:2015ava}
\bea
S_{\rm f}^{LLL}({k})=-i~e^{-{{\bf k}_\bot^2\over|q_{\rm f}B|}}{m\!-\!k_4\gamma^4\!-\!k_3\gamma^3\over k_4^2\!+\!k_3^2\!+\!m^2}[1\!+\!{\rm sgn}(q_{\rm f}B)i\gamma^1\gamma^2].\nonumber\\
\eea
Then, after substituting them into Eq.\eqref{SprpP} and Eq.\eqref{VprpP}, the explicit form of the effective inverse propagators of $\pi^0$ and $\bar{\rho}^+_1$ are respectively
\begin{widetext}
\bea
{D}^{-1}_{\pi^0\pi^0}(p)&=&-{1\over2G_S}+N_c\sum_{\rm f=u,d}\int{\di^4k\over (2\pi)^4}{\rm tr}~S_{\rm f}^{LLL}({k}+p)i\gamma^5S_{\rm f}^{LLL}({k})i\gamma^5\nonumber\\
&=&-{1\over2G_S}+8N_c\sum_{\rm f=u,d}\int{\di^4k\over (2\pi)^4}{e^{-{{\bf k}_\bot^2+({\bf k}_\bot+{\bf p}_\bot)^2\over|q_{\rm f}B|}}[m^2\!+\!k_4(k_4+p_4)\!+\!k_3(k_3+p_3)]\over (k_4^2\!+\!k_3^2\!+\!m^2)[(k_4+p_4)^2\!+\!(k_3+p_3)^2\!+\!m^2]},\label{NPLLL}\\
{D}^{-1}_{\bar{\rho}^+_1\bar{\rho}^+_1}(p)&=&-{1\over2G_V}+{2N_c}\int{\di^4k\over (2\pi)^4}{\rm tr}~S_{\rm d}^{LLL}({k}+p)\Gamma_{\bar{\rho}^-_1}S_{\rm u}^{LLL}({k})\Gamma_{\bar{\rho}^+_1}\nonumber\\
&=&-{1\over2G_V}+32N_c\int{\di^4k\over (2\pi)^4}{e^{-{{\bf k}_\bot^2\over|q_{\rm u}B|}-{({\bf k}_\bot+{\bf p}_\bot)^2\over|q_{\rm d}B|}}[m^2\!+\!k_4(k_4+p_4)\!+\!k_3(k_3+p_3)]\over (k_4^2\!+\!k_3^2\!+\!m^2)[(k_4+p_4)^2\!+\!(k_3+p_3)^2\!+\!m^2]}\label{RPLLL}
\eea
\end{widetext}
with the trace "${\rm tr}$" only over Dirac spinor space. If we assume $q_{\rm u}=-q_{\rm d}$, we can immediately recognize the equality of the second terms in Eq.(\ref{NPLLL}) and Eq.(\ref{RPLLL}) up to a factor $2$, which implies the similarity between the magnetic effects to $\pi^0$ and $\bar{\rho}^+_1$. In vanishing energy-momentum limit $p\rightarrow0$, by integrating out the transverse momenta ${\bf k}_\bot$ and inserting the realistic values of $q_{\rm u}$ and $q_{\rm d}$, the effective inverse propagators can be simply reduced to
\bea
-{D}^{-1}_{\pi^0\pi^0}(0)
&=&{1\over2G_S}-{N_c\over \pi}\int{\di^2k\over (2\pi)^2}{|eB|\over k^2+m^2},\label{qcp}\\
-{D}^{-1}_{\bar{\rho}^+_1\bar{\rho}^+_1}(0)
&=&{1\over2G_V}-{16N_c\over9\pi}\int{\di^2k\over (2\pi)^2}{|eB|\over k^2+m^2}.\label{qcr}
\eea
Actually, theys are just the quadratic Ginzburg-Landau (GL) expansion coefficients (QGLECs) around small order parameters $\langle\pi^0\rangle$ and $\langle\bar{\rho}^+_1\rangle$, refer to that around $\langle\pi^\pm\rangle$ in Ref.~\cite{Cao:2015xja}. Note that only the qualitative respondences to the magnetic field effect should be taken seriously here, because $B$ independent loop contributions are not included in all the formulas, that is, the second terms vanish in the limit $B\rightarrow0$. In this respect, even without introducing any explicit regularization scheme which does not change the signs of the divergent terms, some significant qualitative conclusions can already be drawn:
\begin{itemize}
\item[(1)] In the relatively weak magnetic field region where dynamical quark mass is almost $B$-independent, the QGLECs both decrease with magnetic field, thus seem to favor decreasing of meson masses in order to maintain $D^{-1}(p_0,{\bf p}=0)=0$, see Ref.~\cite{Chernodub:2011mc,Liu:2014uwa,Wang:2017vtn,Mao:2018dqe,Liu:2018zag,Avancini:2016fgq,Shushpanov:1997sf}.
\item[(2)] $\bar{\rho}^+_1$ meson responds more strongly than $\pi^0$ meson to magnetic field as the coefficient in front of $B$ in Eq.\eqref{qcr} is larger than that in Eq.\eqref{qcp}, see the steeper $\bar{\rho}^+_1$ mass reduction in Ref.~\cite{Liu:2014uwa,Liu:2018zag,Bali:2017ian}.	
\item[(3)] In LLL approximation, the gap equation becomes
\bea
{m-m_0\over2G_S}-m{N_c\over \pi}\int{\di^2k\over (2\pi)^2}{|eB|\over k^2+m^2}=0,
\eea
from which the MCE can be told directly, see Ref.~\cite{Wang:2017pje} for more detailed discussions on the large $B$ limit. Then, $-{D}^{-1}_{\pi^0\pi^0}(0)={m_0\over2mG_S}$ is positive-definite and thus disfavors NPSF as verified in Ref.~\cite{Liu:2018zag}; $-{D}^{-1}_{\bar{\rho}^+_1\bar{\rho}^+_1}(0)={1\over2G_V}-{16\over18G_S}+{16m_0\over18mG_S}\approx{1\over2G_V}-{16\over18G_S}$ is negative-definite for the chosen model parameters and thus favors VSC. However, if we recover the $B$ independent loop contribution, $-{D}^{-1}_{\bar{\rho}^+_1\bar{\rho}^+_1}(0)$ will be positive in small $B$ region as should be to maintain finite mass there, see Ref.~\cite{Chernodub:2011mc,Liu:2014uwa,Bali:2017ian} and Fig.~\ref{coefficeint} in Sec.\ref{FLL}.
\item[(4)] Due to their different charges of $u$ and $d$ quarks, the magnetic field will definitely induce splitting MCE in principle~\cite{Wang:2018gmj}, which should be taken care of in more realistic three-flavor NJL model~\cite{Klevansky:1992qe,Ferreira:2013tba}. For $\pi^0$ meson, $u$ and $d$ quarks contribute separately through pure flavor polarization loops; while for $\bar{\rho}^+_1$ meson, they contribute through a flavor-mixed polarization loop. Thus, the splitting MCE is expected to have larger consequence on $\bar{\rho}^+_1$ mass than on $\pi^0$ mass, which must be carefully checked before any conclusion is drawn on whether VSC can happen or not.
\end{itemize}

\subsection{Full Landau levels formalism and numerical results}\label{FLL}

By substituting the full Landau levels forms of quark propagators Eq.\eqref{SE} into Eq.\eqref{SprpP} and Eq.\eqref{VprpP}, the effective inverse propagators of $\pi^0$ and $\bar{\rho}^+_1$ are respectively
\begin{widetext}
\bea
-{D}^{-1}_{\pi^0\pi^0}(p)
&=&{1\over2G_S}-4N_c\sum_{\rm f=u,d}\int{\di^4k\over (2\pi)^4}\int{\di s \di s'}e^{-s\left[m^2+(k_4+p_4)^2+(k_3+p_3)^2+({\bf k+p})_\bot^2 {\tanh B_{\rm f}^s\over B_{\rm f}^s}\right]}e^{-s'\left[m^2+k_4^2+k_3^2+{\bf k}_\bot^2 {\tanh B_{\rm f}^{s'}\over B_{\rm f}^{s'}}\right]}\nonumber\\
&&\left[(m^2\!+\!({\bf k}_\parallel\!+\!{\bf p}_\parallel)\cdot{\bf k}_\parallel)(1\!+\!\tanh B_{\rm f}^s\tanh B_{\rm f}^{s'})\!+\!({\bf k}_\bot\!+\!{\bf p}_\bot)\cdot{\bf k}_\bot(1\!-\!\tanh^2 B_{\rm f}^s)(1\!-\!\tanh^2 B_{\rm f}^{s'})\right],\label{NPFLL}\\
-{D}^{-1}_{\bar{\rho}^+_1\bar{\rho}^+_1}(p)
&=&{1\over2G_V}-8N_c\int{\di^4k\over (2\pi)^4}\int{\di s \di s'}e^{-s\left[m^2+(k_4+p_4)^2+(k_3+p_3)^2+({\bf k+p})_\bot^2 {\tanh B_{\rm u}^s\over B_{\rm u}^s}\right]}e^{-s'\left[m^2+k_4^2+k_3^2+{\bf k}_\bot^2 {\tanh B_{\rm d}^{s'}\over B_{\rm d}^{s'}}\right]}\nonumber\\
&&(m^2\!+\!({\bf k}_\parallel\!+\!{\bf p}_\parallel)\cdot{\bf k}_\parallel)(1\!+\!\tanh B_{\rm u}^s)(1-\tanh B_{\rm d}^{s'}).\label{RPFLL}
	\eea
with ${\bf k}_\parallel=(k_4,k_3)$. The LLL results Eq.\eqref{NPLLL} and Eq.\eqref{RPLLL} can be obtained from these expressions in large $B$ limit, which indicates $\tanh B_{\rm u}^s\rightarrow1$ and $\tanh B_{\rm d}^{s'}\rightarrow-1$ due to their different signs of $q_{\rm u}$ and $q_{\rm d}$.
For vanishing three-momentum ${\bf p}={\bf 0}$, they are reduced to the following forms
\bea
-{D}^{-1}_{\pi^0\pi^0}(p_4)
&=&{1\over2G_S}-{N_c}\sum_{\rm f=u,d}{q_{\rm f}B\over4\pi^2}\int{\di s \di s'\over s\!+\!s'}e^{-(s+s')m^2-{ss'\over s+s'}p_4^2}\left[\left(m^2\!+\!{1\over s\!+\!s'}-{ss'\over (s\!+\!s')^2}p_4^2\right)\coth B_{\rm f}^{s+s'}\!+\!{q_{\rm f}B\over\sinh^2B_{\rm f}^{s+s'}}\right]\nonumber\\
&=&{1\over2G_S}-{N_c\over8\pi^2}\sum_{\rm f=u,d}\int{\di s}\int_{-1}^1 {\di u}~e^{-s\left(m^2+{{1-u^2\over4}}p_4^2\right)}\left[\left(m^2\!+\!{1\over s}-{{1-u^2\over4}}p_4^2\right){q_{\rm f}B\over\tanh B_{\rm f}^{s}}\!+\!{(q_{\rm f}B)^2\over\sinh^2B_{\rm f}^{s}}\right],\label{NPFLL1}\\
-{D}^{-1}_{\bar{\rho}^+_1\bar{\rho}^+_1}(p_4)&=&{1\over2G_V}-{N_c\over2\pi^2}\int{\di s \di s'\over s\!+\!s'}{e^{-(s+s')m^2-{ss'\over s+s'}p_4^2}}\left(m^2\!+\!{1\over s\!+\!s'}-{ss'\over (s\!+\!s')^2}p_4^2\right){(1\!+\!\tanh B_{\rm u}^s)(1-\tanh B_{\rm d}^{s'})\over s{\tanh B_{\rm u}^s\over B_{\rm u}^s}+s'{\tanh B_{\rm d}^{s'}\over B_{\rm d}^{s'}}}\nonumber\\
&=&{1\over2G_V}-{N_c\over4\pi^2}\int\!{\di s\over s}\int_{-1}^1\! {\di u}~{e^{-s\left(m^2+u^+u^-p_4^2\right)}}\left(m^2\!+\!{1\over s}-u^+u^-p_4^2\right){\left[1\!+\!\tanh{B_{\rm u}^s}^+\right]\!\!\left[1\!-\!\tanh{B_{\rm d}^s}^-\right]\over {\tanh{B_{\rm u}^s}^+/B_{\rm u}^s}+{\tanh{B_{\rm d}^s}^-/B_{\rm d}^{s}}}\label{RPFLL1}
\eea
\end{widetext}
by integrating out the internal energy-momentum, where $u^\pm={1\pm u\over2}, {B_{\rm u}^s}^+={B_{\rm u}^s}u^+$ and ${B_{\rm d}^s}^-={B_{\rm d}^s}u^-$ for brevity. 

Then, the effective inverse propagator of $\pi^0$ can be regularized by adopting vacuum regularization scheme as~\cite{Cao:2014uva}	
\bea
-{D}^{-1}_{\pi^0\pi^0}
&=&{1\over2G_S}+\Delta\Pi_{\pi^0\pi^0}-8N_c\int^{\rm reg}{\di^4k\over (2\pi)^4}\nonumber\\
&&{k_4(k_4+p_4)\!+\!E_{\bf k}^2\over (k_4^2\!+\!E_{\bf k}^2)[(k_4+p_4)^2\!+\!E_{\bf k}^2]}\label{NPFLLr}
\eea
with $\Delta\Pi_{\pi^0\pi^0}(p_4)=\Pi_{\pi^0\pi^0}(p_4)-(B\rightarrow0)$ and the quark dispersion $E_{\bf k}=\sqrt{m^2+{\bf k}^2}$.. The story with vector $\bar{\rho}^+_1$ meson is not so simple, because the divergence associated with the $S_z-B$ coupling, the terms odd on $B$ in Eq.\eqref{RPFLL1}, can not be canceled out by any term in vanishing $B$ limit. Expanding the polarization loop in Eq.\eqref{RPFLL1} to linear term on $B$, we have
\bea
\Pi_{\bar{\rho}^+_1\bar{\rho}^+_1}^{o(B^2)}(p_4)&=&-{N_c\over4\pi^2}\int{\di s\over s}\int_{-1}^1 {\di u}~{e^{-s\left(m^2+{{1-u^2\over4}}p_4^2\right)}}\nonumber\\
&&\left(m^2\!+\!{1\over s}-{{1-u^2\over4}}p_4^2\right)\left(1+{eBs\over 2}\right).
\eea
The un-regularized coefficients for zeroth and first orders of $B$ can be put in alternative energy-momentum integration forms
by recognizing the corresponding terms in Eq.\eqref{RPFLL} and integrating out proper time first. Finally, we're ready to perform a modified vacuum regularization to the effective inverse propagator of $\bar{\rho}^+_1$ meson and get
\bea\label{RPFLLr}
-{D}^{-1}_{\bar{\rho}^+_1\bar{\rho}^+_1}
&=&{1\over2G_V}\!+\!\Delta\Pi_{\bar{\rho}^+_1\bar{\rho}^+_1}\!-\!8N_c\!\int^{\rm reg}\!\!\!\!{\di^4k\over (2\pi)^4}\!\!\left(1\!+\!{eB\over k_4^2\!+\!E_{\bf k}^2}\right)\nonumber\\
&&{m^2\!+\!k_4(k_4+p_4)\!+\!k_3^2\over (k_4^2\!+\!E_{\bf k}^2)[(k_4+p_4)^2\!+\!E_{\bf k}^2]}
\eea
with $\Delta\Pi_{\bar{\rho}^+_1\bar{\rho}^+_1}(p_4)=\Pi_{\bar{\rho}^+_1\bar{\rho}^+_1}(p_4)-\Pi_{\bar{\rho}^+_1\bar{\rho}^+_1}^{o(B^2)}(p_4)$.

The great advantage of vacuum regularization is that there are no artifacts for the $B$-dependent parts even when $|eB|^{1/2}$ is much larger than the effective cutoff $\Lambda$ induced by the regularization. This is obvious for $\pi^0$ meson because there is no cutoff in the $B$-dependent term $\Delta\Pi_{\pi^0\pi^0}(p_4)$; for $\bar{\rho}^+_1$ meson, $\Lambda$ is involved in the $B$-linear coefficient but this is just total spin regularization and has nothing to do with $B$. Before analytically continuing the results Eq.\eqref{NPFLLr} and Eq.\eqref{RPFLLr} to Minkovski space through $p_4\rightarrow i p_0$ in order to explore meson spectra, one should remember that the proper-time integration should always be carried out first to give algebraic functions of $p_4$ in principle. Otherwise, ultraviolet divergences will be encountered in the integrations for $p_0\geq2m$ even though it is still fine for $p_0<2m$. The proper-time integration can be gotten rid of directly by adopting the Landau-level presentations of quark propagators. Then, the inverse meson propagators would depend on two series of Landau-level summations and no severe divergences happen for $p_0\geq2m$ anymore. Nevertheless, the compacter proper-time presentation can still be adjusted to suit such explorations through the variable transformation $s\left(m^2+{{1-u^2\over4}}p_4^2\right)\rightarrow s$. In Appendix.~\ref{equality}, we show the equality between proper-time and Landau-level presentations quantitatively up to $p_0=2m$. 

Even though we can handle the potential artificial divergence through mathematic approaches, the case $p_0\geq2m$ still induces nonphysical consequences. In Appendix.~\ref{invalidity}, we compare several regularization schemes at vanishing magnetic field and show the invalidity of NJL model to study the magnetic field effect to $\bar{\rho}^+_1$ meson with physical mass. The main reason is the lack of confinement effect in the NJL model. And even the extensive Polyakov--Nambu--Jona-Lasinio (PNJL) model cannot help the situation, because the thermodynamic potential for the quark part is the same as that in the NJL model at zero temperature~\cite{Ferreira:2013tba}. For the purpose of qualitative study, we assume $\rho$ meson mass to be a bit smaller than twice quark mass $2m$ in the vacuum and use three-momentum cutoff scheme to regularize the vacuum terms. Then, the effective inverse meson propagators are explicitly
\bea
\!\!\!\!\!\!\!-{D}^{-1}_{\pi^0\pi^0}
={1\over2G_S}+\Delta\Pi_{\pi^0\pi^0}-N_c\int_0^\Lambda {k^2dk\over\pi^2}\frac{8E_k}{4 E_k^2+p_4^2},\label{NPFLLr1}
\eea
\bea
-{D}^{-1}_{\bar{\rho}^+_1\bar{\rho}^+_1}
\!&=&\!{1\over2G_V}+\Delta\Pi_{\bar{\rho}^+_1\bar{\rho}^+_1}-N_c\int_0^\Lambda {k^2dk\over \pi^2E_k}\left[\frac{8(m^2\!+\!{2\over3}k^2)}{4 E_k^2+p_4^2}\right.\nonumber\\
&&\left.+\frac{8E_k^2({2m^2\!+\!k^2})\!-\!{2\over3}k^2p_4^2}{E_k^2(4 E_k^2+p_4^2)^2}eB\right]
\label{RPFLLr1}
\eea
by carrying out the integration over energy $k_0$, which are consistent with the corresponding ones given in Ref.\cite{Klevansky:1992qe,Wang:2018gmj,Brauner:2016lkh} for vanishing $B$ case. 
\begin{figure}[!htb]
	\begin{center}
		\includegraphics[width=8cm]{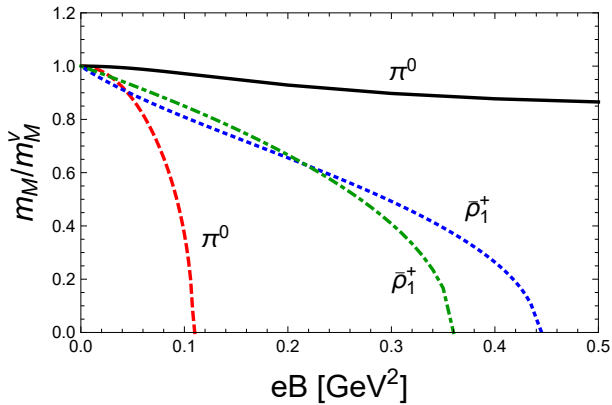}
		\caption{(color online) The self-consistent masses of $\pi^0$ (black solid line) and $\bar{\rho}^+_1$ (blue dotted line) mesons as functions of magnetic field $B$ in two-flavor NJL model. For comparison, $\pi^0$ mass with fixed quark mass (red dashed line) and point particle mass for $\bar{\rho}^+_1$ (green dot-dashed line) are also included. All meson masses $m_M$ are normalized to their vacuum masses $m_M^v$.}\label{mass_2f}
	\end{center}
\end{figure}

\begin{figure}[!htb]
	\begin{center}
		\includegraphics[width=8cm]{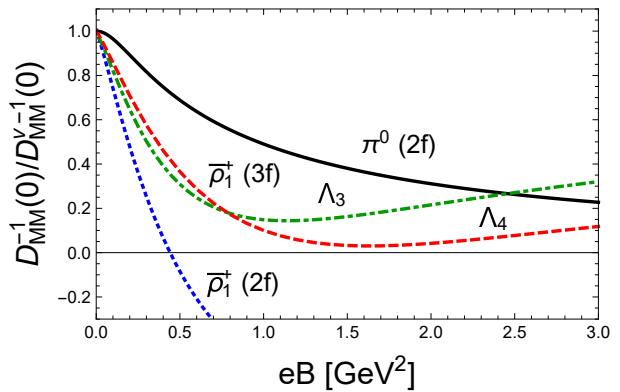}
		\caption{(color online) The quadratic GL expansion coefficients as functions of $B$ for $\pi^0$ (black solid line) and $\bar{\rho}^+_1$ (blue dotted line) mesons, respectively, in two-flavor NJL model. The ones for $\bar{\rho}^+_1$ in three-flavor NJL model with four-momentum cutoff ($\Lambda_4$) parameterization~\cite{Klimt:1989pm} (red dashed line) and three-momentum cutoff ($\Lambda_3$) parameterization ~~\cite{Rehberg:1995kh} (green dot-dashed line) are also shown for comparison. They are all normalized to their vacuum values $-D_{MM}^{v\ -1}~(>0)$.}\label{coefficeint}
	\end{center}
\end{figure}

Armed with the regularized gap equation Eq.\eqref{gapm} and effective inverse meson propagators Eq.\eqref{NPFLLr1} and Eq.\eqref{RPFLLr1}, we are ready to perform further numerical calculations. In two-flavor NJL model, the model parameters are fixed as $G_V=3.37~{\rm GeV}^{-2}$, $G_S=4.93~{\rm GeV}^{-2}$, $\Lambda=0.653~{\rm GeV}$ and $m_0=5~{\rm MeV}$ by fitting to rho meson mass $m_\rho^v=0.6~{\rm GeV}$ (smaller than $2m_v=0.626~{\rm GeV}$), pion mass $m_\pi^v=0.134~{\rm GeV}$, pion decay constant $f_\pi=93~{\rm MeV}$ and quark condensate $\langle\sigma\rangle=-2\times (0.25~{\rm GeV})^3$ in vacuum ~\cite{Zhuang:1994dw}. The self-consistent meson mass spectra and QGLECs are illuminated in Fig.\ref{mass_2f} and Fig.\ref{coefficeint}, respectively. For comparison, the results with fixed quark mass for $\pi^0$ meson and point particle mass $\sqrt{{m_\rho^v}^2-|eB|}$ for $\bar{\rho}^+_1$ meson are also demonstrated as functions of magnetic field in Fig.\ref{mass_2f}. 

As we can see in Fig.\ref{mass_2f}, if the MCE is suppressed, $\pi^0$ mass will quickly decrease to zero thus favors NPSF. The reality is that $\pi^0$ mass only mildly decreases with $B$ in the weak field region and slowly increases in the strong field region (the scale is not shown here), which is consistent with the previous NJL model result~\cite{Liu:2018zag,Avancini:2016fgq}. So it's the consistent gap equation that forbids the mass of the pseudo-Goldstone boson $\pi^0$ to decrease to zero. The self-consistent $\bar{\rho}^+_1$ mass decreases to zero at some point thus favors VSC, which is consistent with the previous two-flavor NJL model exploration~\cite{Liu:2014uwa}. However, contrary to the advance of VSC compared to the point particle pattern  in Ref.\cite{Liu:2014uwa} while with weak $B$ approximation, the delay of VSC due to the MCE to the composite quark mass in present calculations justifies our more careful explorations in the three-flavor NJL model in Sec.\ref{3flavor}. Finally, the quadratic coefficients in Fig.\ref{coefficeint} indicate much stronger respondence of $\bar{\rho}^+_1$ to magnetic field than $\pi^0$, which agrees with the mass spectra in Fig.\ref{mass_2f} and the qualitative discussions in Sec.\ref{LLL}.

\section{Meson spectra within three-flavor Nambu--Jona-Lasinio model}\label{3flavor}
As pointed out in the end of Sec.\ref{LLL}, the magnetic field inevitably induces splitting MCE to $u$ and $d$ quark masses, which requires the iso-vector scalar interaction channels in the model for mean field exploration. In this respect, the three-flavor NJL model is much more suitable for realistic study, the Lagrangian density of which can be extended from the previous one to~\cite{Klevansky:1992qe,Klimt:1989pm}
\begin{eqnarray}
{\cal L}_{\rm NJL}&=&\bar\psi(i\slashed{D}-m_0)\psi+G_S\sum_{a=0}^8[(\bar\psi\lambda^a\psi)^2+(\bar\psi i\gamma_5\lambda^a\psi)^2]\nonumber\\
&&+{\cal L}_6-G_V\left[\left(\bar\psi\gamma^\mu\tau^a\psi\right)^2+\left(\bar\psi i\gamma^\mu\gamma_5\tau^a\psi\right)^2\right]\nonumber\\
&{\cal L}_6&=-K\sum_{s=\pm}{\rm Det}\bar\psi\Gamma^s\psi
\end{eqnarray} 
by further adopting the four fermion vector interaction channels with coupling constant $G_V$. Compared to the two-flavor case, $\psi=(u,d,s)^T$ represents the three-flavor quark field, $m_0={\rm diag}(m_{\rm 0u},m_{\rm 0d},m_{\rm 0s})$ is the current quark mass matrix, and the covariant derivative is defined as $D_\mu=\partial_\mu-iQA_\mu$ with the charge matrix $Q={\rm diag}(q_{\rm u},q_{\rm d},q_{\rm s})$. For the four-fermion interaction terms, $\lambda^0=\sqrt{2\over3}I$ and Gell-Mann matrices $\lambda^i~(i=1,\dots,8)$ are defined in three-flavor space, so the extra diagonal terms $(\bar\psi\lambda^{3}\psi)^2$ and $(\bar\psi\lambda^{8}\psi)^2$ allow mass splitting among all the flavors compared to the two-flavor case. The $U_A(1)$ symmetry violating six-fermion interactions~\cite{tHooft:1976snw} only involve scalar-pseudoscalar channels with the determinant defined in flavor space, $\Gamma^\pm=1\pm\gamma_5$ and $K$ the coupling constant. Now, following the same ansatz as the two-flavor case, we only consider nonzero chiral condensations $\sigma_{\rm i}\equiv\langle\bar{\psi}^i{\psi}^i\rangle$ with $i$ flavor index\footnote{Here and in the following, the correspondence between the number index $i=1,2,3$ and the more explicit Latin index ${\rm f=u,d,s}$ should be understood. We prefer the explicit Latin presentation whenever convenient.} and the $U_A(1)$ symmetry violating term ${\cal L}_6$ can be reduced to an effective four fermion interaction form in Hartree approximation~\cite{Klevansky:1992qe}:
\begin{widetext}
\bea
{\cal L}_6^4&=&-{K\over2}\sum_{s=\pm}\epsilon_{ijk}\epsilon_{imn}\langle\bar{\psi}^i\Gamma^s{\psi}^i\rangle(\bar{\psi}^j\Gamma^s{\psi}^m)(\bar{\psi}^k\Gamma^s{\psi}^n)\nonumber\\
&=&-{K\over6}\Big\{2\sum_{\rm f=u,d,s}\sigma_{\rm f}(\bar{\psi}\lambda^0\psi)^2-3\sigma_s\sum_{i=1}^3(\bar{\psi}\lambda^i\psi)^2
-3\sigma_{\rm d}\sum_{i=4}^5(\bar{\psi}\lambda^i\psi)^2-3\sigma_{\rm u}\sum_{i=6}^7(\bar{\psi}\lambda^i\psi)^2+(\sigma_s\!-\!2\sigma_{\rm u}\!-\!2\sigma_{\rm d})(\bar{\psi}\lambda^8\psi)^2\nonumber\\
&&+\sqrt{2}(2\sigma_s\!-\!\sigma_{\rm u}\!-\!\sigma_{\rm d})(\bar{\psi}\lambda^0\psi)(\bar{\psi}\lambda^8\psi)-\sqrt{6}(\sigma_{\rm u}\!-\!\sigma_{\rm d})(\bar{\psi}\lambda^3\psi)(\bar{\psi}\lambda^0\psi-\sqrt{2}\bar{\psi}\lambda^8\psi)\Big\}-(\lambda^a\rightarrow i\lambda^a\gamma^5)
\eea
with $\epsilon_{ijk}$ the Levi-Civita symbol. So the reduced three-flavor Lagrangian density with only four fermion effective interactions is
\begin{eqnarray}
\!\!\!{\cal L}_{\rm NJL}^4=\bar\psi(i\slashed{D}-m_0)\psi+\!\!\sum_{a,b=0}^8\!\left[G_{ab}^-(\bar\psi\lambda^a\psi)(\bar\psi\lambda^b\psi)\!+\!G_{ab}^+(\bar\psi i\gamma_5\lambda^a\psi)(\bar\psi i\gamma_5\lambda^b\psi)\right]\!-\!G_V\!\!\left[\left(\bar\psi\gamma^\mu\tau^a\psi\right)^2\!+\!\left(\bar\psi i\gamma^\mu\gamma_5\tau^a\psi\right)^2\right],
\end{eqnarray}
where the non-vanishing elements of the symmetric coupling matrices $G^\pm$ are given by~\cite{Klevansky:1992qe}
\begin{eqnarray}
&&G_{00}^\mp=G_S\mp {K\over3}\sum_{\rm f=u,d,s}\sigma_{\rm f},~G_{11}^\mp=G_{22}^\mp=G_{33}^\mp=G_S\pm {K\over2}\sigma_s,~G_{44}^\mp=G_{55}^\mp=G_S\pm {K\over2}\sigma_{\rm d},~G_{66}^\mp=G_{77}^\mp=G_S\pm {K\over2}\sigma_{\rm u},\nonumber\\
&&G_{88}^\mp=G_S\mp {K\over6}(\sigma_s-2\sigma_{\rm u}-2\sigma_{\rm d}),~G_{08}^\mp=\mp {\sqrt{2}K\over12}(2\sigma_s\!-\!\sigma_{\rm u}\!-\!\sigma_{\rm d}),~G_{38}^\mp=-\sqrt{2}G_{03}^\mp=\mp {\sqrt{3}K\over6}(\sigma_{\rm u}\!-\!\sigma_{\rm d}).
\end{eqnarray}

By contracting a pair of field and conjugate field operators further in ${\cal L}_6^4$ in Hartree approximation, we find
\begin{eqnarray}
{\cal L}_6^2&=&-\sum_{s=\pm}^{i(\neq j\neq k)}K\langle\bar{\psi}^j\Gamma^s{\psi}^j\rangle\langle\bar{\psi}^k\Gamma^s{\psi}^k\rangle[\bar{\psi}^i\Gamma^s{\psi}^i]=-K\sum_{ijk}\!\epsilon_{ijk}^2\bar{\psi}^i\sigma_j\sigma_k{\psi}^i,
\end{eqnarray}
which then, together with the contributions from the initial four-quark interactions, gives the effective quark masses as
\begin{eqnarray}\label{massi}
m_i^*&=&m_{0i}-4G_S\sigma_i+K\sum_{jk}\!\epsilon_{ijk}^2\sigma_j\sigma_k.
\end{eqnarray}
In order to evaluate quark masses numerically, we should be equipped with the gap equations directly following the definitions of chiral condensations:
\begin{eqnarray}
\sigma_i\equiv\langle\bar{\psi}^i{\psi}^i\rangle=-{i\over V_4}{\rm Tr}~G_{i},
\end{eqnarray}
where the effective quark propagators in a constant magnetic field, $G_{i}$, can be modified from Eq.\eqref{qprp} by just altering $m$ to $m_i$ for different flavors. Then, by following similar derivations and discussions as in the two-flavor case, the regularized gap equations are
\begin{eqnarray}
-\sigma_{\rm f}
&=&N_c{{m_{\rm f}^*}^3\over2\pi^2}\Big[\tilde{\Lambda}_{\rm f}\Big({1+\tilde{\Lambda}_{\rm f}^2}\Big)^{1\over2}-\ln\Big({\tilde\Lambda_{\rm f}}
+\Big({1+\tilde{\Lambda}_{\rm f}^2}\Big)^{1\over2}\Big)\Big]+N_c{m_{\rm f}^*\over4\pi^2}\int_0^\infty {ds\over s^2}e^{-{m_{\rm f}^*}^2s}\left({q_{\rm f}Bs
	\over\tanh(q_{\rm f}Bs)}-1\right).\label{mgap}
\end{eqnarray}
with the reduced cutoff $\tilde{\Lambda}_{\rm f}={\Lambda/m_{\rm f}^*}$.
In advance, the thermodynamic potential can be obtained consistently by combining the definitions of effective masses in Eq.\eqref{massi} and the integrations over ${\sigma_{\rm f}}$ of Eq.\eqref{mgap}:
	\begin{eqnarray}
	\Omega&=&2G_S\sum_{{\rm f}=u,d,s}\sigma_{\rm f}^2-4K\prod_{{\rm f}=u,d,s}\sigma_{\rm f}-N_c\sum_{{\rm f}=u,d,s}\left\{{{m_{\rm f}^*}^4\over8\pi^2}\Big[\tilde{\Lambda}_{\rm f}\Big(1+{2\tilde{\Lambda}_{\rm f}^2}\Big)\Big({1+{\tilde{\Lambda}_{\rm f}^2}}\Big)^{1\over2}-\ln\Big({\tilde{\Lambda}_{\rm f}}
+\Big({1+{\tilde{\Lambda}_{\rm f}^2}}\Big)^{1\over2}\Big)\Big]\right.\nonumber\\
&&\left.-{1\over8\pi^2}\int_0^\infty {ds\over s^3}e^{-{m_{\rm f}^*}^2s}\left({q_{\rm f}Bs
	\over\tanh(q_{\rm f}Bs)}-1\right)\right\},
\end{eqnarray}
which is consistent with that in Ref.~\cite{Ferreira:2013tba}.

Finally, let's focus on the collective excitation modes, especially the neutral pseudoscalar and vector modes. It is helpful to define the one-flavor polarization loop according to Eq.\eqref{SprpP}:
	\begin{eqnarray}
	\Pi_{\rm f}&=&-N_c\int{\di^4k\over (2\pi)^4}\text{tr}~S_{\rm f}({k}+p)i\gamma^5S_{\rm f}({k})i\gamma^5,
	\end{eqnarray}
which can be regularized in the same way as the two-flavor case.
Then, by setting the three-dimensional diagonal matrix $\Pi^+_0\equiv{\rm diag}(\Pi_{\rm u},\Pi_{\rm d},\Pi_{\rm s})$,
the polarization functions in the neutral pseudoscalar sector, $\Pi^{+}_{ij}\equiv-{\rm Tr}S({k}+p)i\gamma^5\lambda^iS({k})i\gamma^5\lambda^j$ with $i,j=0,3,8$\footnote{Note that the subscripts $0,3$ and $8$ correspond to neutral pseudoscalar $\eta_0, \pi^0$ and $\eta_8$ channels, respectively.}, can be evaluated directly through $\Pi^{+}_{ij}={\rm tr_f}~\lambda^i\Pi^+_0\lambda^j$. As $\Pi^+_0$ and $\lambda^i$ are all diagonal matrices, $\Pi^{+}_{ij}$ is symmetric with respect to interchange of the subscripts $i$ and $j$ and only $6$ independent functions are involved. Thus, the effective inverse propagator matrix of the neutral pseudoscalar sector is
	\begin{eqnarray}\label{NPSM}
	-D^{-1}_{ij}(p)={1\over2}({G}^{+})^{-1}_{ij}+\Pi^{+}_{ij},
	\end{eqnarray}
	where the explicit forms of the polarization functions are
	\bea
	&&\Pi^{+}_{00}={2\over3}\sum_{\rm f=u,d,s}\Pi_{\rm f},~\Pi^{+}_{03}=\sqrt{2\over3}(\Pi_{\rm u}-\Pi_{\rm d}),~\Pi^{+}_{08}={\sqrt{2}\over3}(\Pi_{\rm u}+\Pi_{\rm d}-2\Pi_{\rm s}),\nonumber\\
	&&\Pi^{+}_{33}=\Pi_{\rm u}+\Pi_{\rm d},~\Pi^{+}_{38}=\sqrt{1\over3}(\Pi_{\rm u}-\Pi_{\rm d}),~\Pi^{+}_{88}={1\over3}(\Pi_{\rm u}+\Pi_{\rm d}+4\Pi_{\rm s}).
	\eea
Besides the mesonic flavor mixing between $\eta_0$ and $\eta_8$ channels due to $U_A(1)$ anomaly in vacuum~\cite{tHooft:1976snw}, magnetic field develops further mixing among all the channels as all the non-diagonal elements of $\Pi^{+}_{ij}$ are non-vanishing now. For simplicity, the mixing between pseudoscalar and pseudovector sectors is neglected -- this is valid as we find the pseudoscalar masses do not change much compared to those in Ref.~\cite{Klimt:1989pm}. Then, the pole masses of the neutral pseudoscalar mesons can be solved numerically by following the condition $\det D^{-1}_{ij}(p_0,{\bf p}={\bf 0})=0$~\cite{Klimt:1989pm} and three independent solutions can be obtained in principle.
	
For the most interested vector mode $\bar{\rho}^+_1$, the change comes from the possibly different masses between $u$ and $d$ quarks, which alters Eq.\eqref{RPFLL} to
\bea	
-{D}^{-1}_{\bar{\rho}^+_1\bar{\rho}^+_1}(p)&\equiv&{1\over2G_V}+\Pi_{\bar{\rho}^+_1\bar{\rho}^+_1}^*(p_4)\nonumber\\
&=&{1\over2G_V}-8N_c\int{\di^4k\over (2\pi)^4}\int{\di s \di s'}e^{-s\left[{m^*_{\rm u}}^2+(k_4+p_4)^2+(k_3+p_3)^2+({\bf k+p})_\bot^2 {\tanh B_{\rm u}^s\over B_{\rm u}^s}\right]}e^{-s'\left[{m^*_{\rm d}}^2+k_4^2+k_3^2+{\bf k}_\bot^2 {\tanh B_{\rm d}^{s'}\over B_{\rm d}^{s'}}\right]}\nonumber\\
&&(m^*_{\rm u}m^*_{\rm d}\!+\!(k_4\!+\!p_4)k_4\!+\!(k_3\!+\!p_3)k_3)(1\!+\!\tanh B_{\rm u}^s)(1-\tanh B_{\rm d}^{s'})\nonumber\\
&\stackrel{{\bf p}={\bf 0}}{=}&{1\over2G_V}\!-\!{N_c\over4\pi^2}\!\!\int\!{\di s\over s}\!\!\int_{-1}^1\!\! {\di u}~{e^{-s\left[{m^*_{\rm u}}^2u^+\!+{m^*_{\rm d}}^2u^-\!+u^+u^-p_4^2\right]}}\!\!\left(m^*_{\rm u}m^*_{\rm d}\!+\!{1\over s}-u^+u^-p_4^2\right)\!\!{\left[1\!+\!\tanh{B_{\rm u}^s}^+\right]\!\!\!\left[1\!-\!\tanh{B_{\rm d}^s}^-\right]\over {\tanh{B_{\rm u}^s}^+\over B_{\rm u}^s}+{\tanh{B_{\rm d}^s}^-\over B_{\rm d}^{s}}}.\nonumber\\
\eea
It can be regularized by following a similar procedure as in Sec.\ref{FLL} and we get
\bea\label{RP3f}	
-{D}^{-1}_{\bar{\rho}^+_1\bar{\rho}^+_1}
&=&{1\over2G_V}+\Delta\Pi_{\bar{\rho}^+_1\bar{\rho}^+_1}^*-N_c\int_0^\Lambda {k^2dk\over\pi^2}\frac{[(E_{\rm u}+E_{\rm d})^2-(m^*_{\rm u}-m^*_{\rm d})^2-{4\over3}k^2](E_{\rm u}+E_{\rm d})}{E_{\rm u}E_{\rm d}[(E_{\rm u}+E_{\rm d})^2+p_4^2]}\nonumber\\
&&-{4N_c}\int_0^\Lambda {k^2dk\over\pi^2}\left\{{q_{\rm u}B}\int_{-\infty}^{\infty}{\di k_4\over2\pi}\frac{{m^*_{\rm u}}{m^*_{\rm d}}+(k_4+p_4)k_4+k_3^2}{[(k_4+p_4)^2+k_3^2+{m^*_{\rm u}}^2]^2(k_4^2+k_3^2+{m^*_{\rm d}}^2)}-(u\leftrightarrow d)\right\}\nonumber\\
&=&{1\over2G_V}+\Delta\Pi_{\bar{\rho}^+_1\bar{\rho}^+_1}^*-N_c\int_0^\Lambda\!\! {2k^2dk\over\pi^2}\frac{(E_{\rm u}E_{\rm d}\!+\!{m^*_{\rm u}}{m^*_{\rm d}}\!+\!{1\over3}k^2)(E_{\rm u}\!+\!E_{\rm d})}{E_{\rm u}E_{\rm d}[(E_{\rm u}\!+\!E_{\rm d})^2\!+\!p_4^2]}-{N_c}\int_0^\Lambda {k^2dk\over\pi^2}\left\{{{q_{\rm u}B}\over(E_{\rm u}\!+\!E_{\rm d})^2\!+\!p_4^2}\right.\nonumber\\
&&\left.\left[\left(\frac{E_{\rm u}E_{\rm d}\!+\!{m^*_{\rm u}}{m^*_{\rm d}}\!+\!{1\over3}k^2}{E_{\rm u}^3}\!+\!{1\over E_{\rm u}}\!+\!{1\over E_{\rm d}}\right)\!-\!\frac{[p_4^2\!+\!(m^*_{\rm u}\!-\!m^*_{\rm d})^2\!+\!{4\over3}k^2](E_{\rm u}\!+\!E_{\rm d})^2}{E_{\rm u}^2E_{\rm d}[(E_{\rm u}+E_{\rm d})^2+p_4^2]}\right]\!-\!(u\leftrightarrow d)\right\},
\eea
where $\Delta\Pi_{\bar{\rho}^+_1\bar{\rho}^+_1}^*(p_4)=\Pi_{\bar{\rho}^+_1\bar{\rho}^+_1}^*(p_4)-\Pi_{\bar{\rho}^+_1\bar{\rho}^+_1}^{o(B^2)*}(p_4)$, the dispersions are $E_{\rm f}=\sqrt{{m_{\rm f}^*}^2+k^2}$ and which reduces exactly to Eq.\eqref{RPFLLr1} if $m_{\rm u}^*=m_{\rm d}^*=m$. The simpler three-momentum cutoff scheme is adopted for the regularization of the divergent terms all through this section. Actually, our numerical calculations with $\rho$ meson vacuum mass $m_\rho^v=0.7~{\rm GeV}$ show that the patterns of QGLECs and thus the ground states do not depend on the choices of regularization schemes, see Fig.\ref{coefficeint}.
\end{widetext}

\begin{figure}[!htb]
	\begin{center}
		\includegraphics[width=8cm]{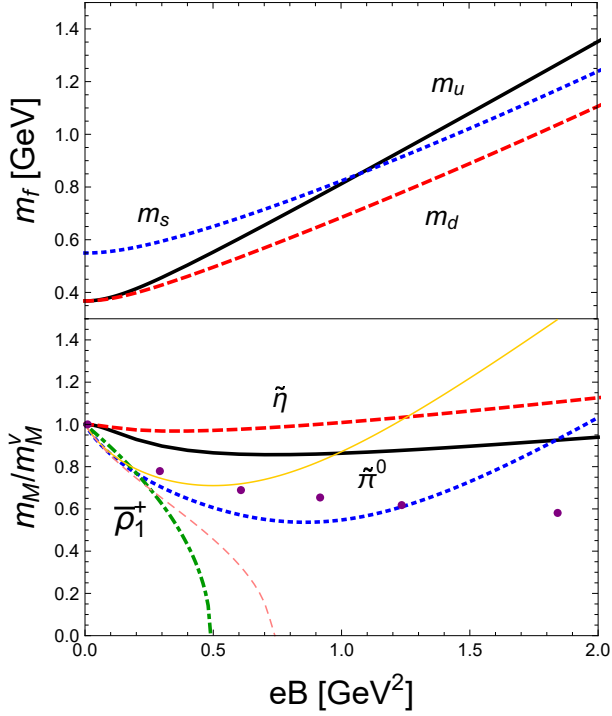}
		\caption{(color online) Upper panel: the evolutions of quark masses with magnetic field $B$ in three-flavor NJL model. Lower panel: the self-consistent masses of $\tilde{\pi}^0$ (black solid line), $\tilde\eta$ (red dashed line) and $\bar{\rho}^+_1$ (blue dotted line) mesons as functions of magnetic field $B$. For comparison, $\bar{\rho}^+_1$ masses from point particle formula (green dot-dashed line) and LQCD simulations (purple points) are also included, which are both adjusted to the vacuum mass $m_\rho^v=0.7~{\rm GeV}$. To help understanding, $\bar{\rho}^+_1$ mass with both composite quark masses equally $m_{\rm u}$ (thin yellow solid line) or  $m_{\rm d}$ (thin pink dashed line) is also shown. All meson masses $m_M$ are normalized to their vacuum masses: $m_{\pi^0}^v=0.134~{\rm GeV},~m_{\eta}^v=0.515~{\rm GeV}$ and $m_{\rho}^v=0.7~{\rm GeV}$.}\label{mass}
	\end{center}
\end{figure}

Now, we are ready to study the properties of collective modes in magnetic field through Eq.\eqref{NPSM} and Eq.\eqref{RP3f} after solving the gap equations Eq.\eqref{mgap} self-consistently. In order to perform numerical calculations, we choose the following parameters for the scalar-pseudoscalar sector: $m_{\rm u}=m_{\rm d}=5.5~{\rm MeV}, m_{\rm s}=140.7~{\rm MeV}, \Lambda=602.3~{\rm MeV}, G_S\Lambda^2=1.835$ and $K\Lambda^5=12.36$~\cite{Rehberg:1995kh}. As the dynamical $u/d$ quark vacuum mass in this case ($0.368~{\rm GeV}$) is larger than that in two-flavor case ($0.313~{\rm GeV}$), the vector coupling constant is fixed to $G_V\Lambda^2=2.527$ by fitting to larger $\rho$ meson vacuum mass $m_\rho^v=0.7~{\rm GeV}$. First of all, the $B$-dependence of the dynamical quark masses $m_{\rm f}$ is illuminated in the upper panel of Fig.\ref{mass}. Three main observations follow: $(1)$ mass splitting between $u$ and $d$ quarks is developed at larger $B$ thus confirming splitting MCE; $(2)$ $m_{\rm u}$ increases most quickly due to its larger electric charge, see also Ref.~\cite{Wang:2018gmj}; $(3)$ $m_{\rm d}$ and $m_{\rm s}$ increase parallelly to each other at larger $B$ due to the same electric charge.

Next, the masses of the interested eigenstates are illuminated in the lower panel of Fig.\ref{mass}, where $\tilde{\pi}^0$ and $\tilde\eta$ are the effective neutral pseudoscalar mesons corresponding to ${\pi}^0$ and $\eta$ meson at vanishing $B$. All the meson masses obtained consistently in the three-flavor NJL model show a similar feature with $B$: first decreasing and then increasing, though the variations of the effective neutral meson masses are much milder than the charged vector meson $\bar{\rho}^+_1$. The latter is consistent with the different constructions of the corresponding polarization loops as discussed in Sec.\ref{LLL} and the enhancement at larger $B$ is due to the domination of splitting MCE among the quarks. To help understand the underlying physics, the $\bar{\rho}^+_1$ mass with both composite quark masses chosen to be equally $m_{\rm u}$ or $m_{\rm d}$ is also demonstrated in the lower panel of Fig.\ref{mass}. As the $\bar{\rho}^+_1$ mass decreases to zero when $m_{\rm d}$ is adopted also for $u$ quark, we can easily conclude that the great enhancement of $m_{\rm u}$ balances the $\bar{\rho}^+_1$ mass reduction at larger $B$. The normalized $\bar{\rho}^+_1$ mass with point particle formula and in LQCD simulations~\cite{Bali:2017ian} are also shown for comparison. Our results are semi-quantitatively consistent with that from the LQCD at relatively weak magnetic field region. There the curvatures are both positive with respect to $B$, in contrary to negative ones from the point particle formula and two-flavor NJL model study (except for the weakest $B$ region). 
 
\begin{figure}[!htb]
	\begin{center}
		\includegraphics[width=8cm]{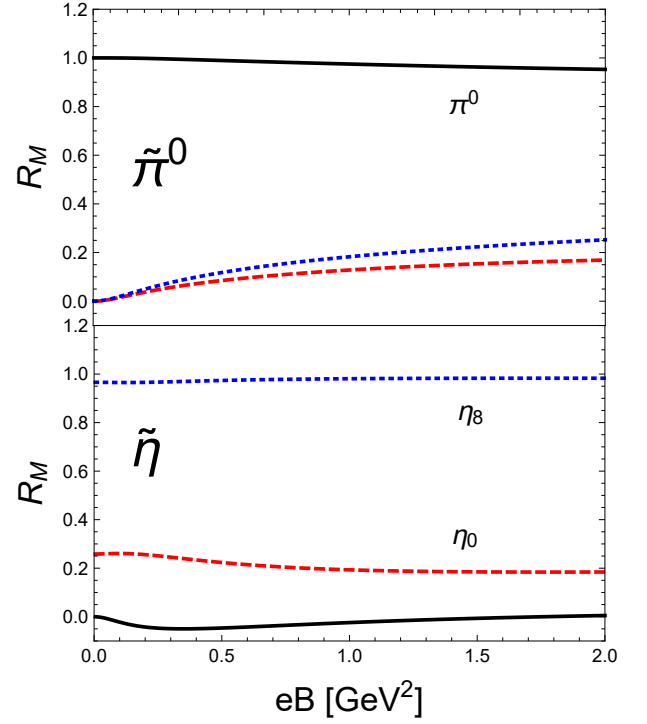}
		\caption{The evolutions of the normalized mixing factors $R_M$ of the effective neutral pseudoscalar $\tilde{\pi}^0$ (upper panel) and $\tilde\eta$ (lower panel) with magnetic field $B$ in terms of $\pi^0$ (black solid line), $\eta_0$ (red dashed line) and $\eta_8$ (blue dotted line).}\label{PSM}
	\end{center}
\end{figure}

Finally, it is intriguing to explore the mixing features of the neutral pseudoscalar mesons with respect to $B$ as we've argued before. The normalized mixing factors $R_M$ of the effective neutral pseudoscalar mesons on their mass shells are shown together in Fig.\ref{PSM} in terms of $\pi^0, \eta_0$ and $\eta_8$. Though, the fractions of $\eta_0$ and $\eta_8$ are small in the whole region, they're very important to keep the effective meson $\tilde{\pi}^0$ light, otherwise ${\pi}^0$ mass will increase to $2.5m_{\pi^0}^v$ at $eB=2~{\rm GeV}^2$. Other interesting observations are that the ratio of the pure flavor component $ \bar{u}i\gamma^5u$ enhances a little in $\tilde{\pi}^0$ and $\tilde\eta\rightarrow \eta_8$ with $B$ increasing, contrary to the naive expectation that $u$ and $d$ quarks will separate from each other quickly in strong magnetic field~\cite{Bali:2017ian}. The reason for the discrepancy is that the effective coupling constants $G^+_{00},G^+_{33}$ and $G^+_{88}$ are quite different from each other and the mixing couplings $G^+_{ij}$ are nonzero for $i\neq j$. In this case, the flavor separation effect, discovered in the $U_A(1)$ symmetric two-flavor NJL model due to the presence of parallel EM field~\cite{Wang:2018gmj}, can never be realized in the three-flavor NJL model at all.

\section{Conclusions and discussions}
In this work, we explored mainly the masses of $\pi^0$ and $\bar{\rho}^+_1$ mesons in external magnetic field, thus the possibility of neutral pion superfluidity and vacuum superconductivity, within the chiral effective two- and three-flavor NJL models. We found similar origins for the reductions of $\pi^0$ and $\bar{\rho}^+_1$ masses in weak $B$ region by adopting lowest Landau level approximation, that is, the linear response coefficients with respect to $B$ are both negative. Because of the magnetic catalysis effect on chiral symmetry breaking or the quark mass, NPSF can never happen in either two- or three-flavor NJL model even with a very strong B, which is consistent with the previous findings~\cite{Bali:2017ian,Mao:2018dqe,Liu:2018zag,Avancini:2016fgq,Hengtong2019}. While the emergence of vacuum superconductivity is delayed compared to point particle result in two-flavor case, it is completely avoided thanks to the splitting MCE among quarks in three-flavor case. It has to be mentioned that choosing a $\bar{\rho}^+_1$ vacuum mass close to the physical value $775~{\rm MeV}$ is also very important to reproduce the LQCD results semi-quantitatively, because Vafa-Witten theorem has no way to play a role in chiral effective models if constrains from real QCD are not well respected. For example, if we set $m_\rho^v=0.6~{\rm GeV}$ in the three-flavor NJL model, VSC will be favored in the medium $B$ region and then disfavored with $B$ increasing further. The discrepancy between our consistent evaluations and LQCD simulations hasn't been well understood yet in larger $B$ region. Even the introduction of asymptotic freedom, which indicates that $G_V$ decreases with $B$, cannot help because $\bar{\rho}^+_1$ mass would enhance further for smaller coupling constant. This is considered to be another {\it puzzle} of QCD in strong magnetic field background. Moreover, mesonic flavor mixing in the neutral pseudoscalar sector is explored in advance, regarding the competition between the $U_A(1)$ anomaly and magnetic field effects: the ratio of pure flavor component $\bar{u}i\gamma^5u$ enhances a little in $\tilde{\pi}^0$ and $\tilde\eta\rightarrow \eta_8$ with increasing $B$. We want to point out that mesonic flavor mixing is very important to keep the masses of the effective eigenstates light, thus the effective neutral pion is still the most relevant degrees of freedom to thermodynamics at very strong $B$. 

Besides the puzzles we proposed in this work, the fates of NPSF and VSC in the presence of parallel rotation and magnetic field are also very interesting topics. With the charged pion superfluidity found and checked in such a system~\cite{Liu:2017spl,CaoandChen}, it is even more convincing that charged rho meson superconductivity can be developed for a sufficiently large rotation due to the meson's more stable $p$-wave spin structure. As a matter of fact, this case does not violate VW theorem, because the rotation itself breaks the positivity of fermion determinant which is a necessary condition for the proof of the theorem, see also the discussions in Ref.~\cite{Cao:2015cka}. Our study suggests the three-flavor NJL model a proper chiral model to explore the magnetic field effect to rho meson properties and thus also a nice candidate for the case with parallel rotation and magnetic field. We suspect that there might be competition between charged pion superfluidity and charged rho superconductivity.

{\bf Acknowledgments:} G.C. appreciates Yoshimasa Hidaka's discussions and comments on this work and is supported by the NSFC Grant No. 11805290.
\appendix
\begin{widetext}
\section{The equality between proper-time and Landau-level presentations}\label{equality}
By solving Dirac equation in external magnetic field, fermion eigenfunctions can be obtained for different Landau levels, from which the effective quark propagators can be constructed as a sum of all Landau level Green's functions~\cite{Miransky:2015ava}. In energy-momentum space, we have
\bea
S_{\rm f}(k)&=&-i~e^{-{{\bf k}_\bot^2\over|q_{\rm f}B|}}\sum_{n=0}^\infty(-1)^n{D_n(q_{\rm f} B,k)\over k_4^2\!+\!k_3^2\!+\!m^2+2n|q_{\rm f} B|},\\
D_n(q_{\rm f} B,k)&=&(m-\slashed k_4-\slashed k_3)\left[{\cal P}_+^{\rm f}L_n\left({2{\bf k}_\bot^2\over|q_{\rm f}B|}\right)-{\cal P}_-^{\rm f}L_{n-1}\left({2{\bf k}_{\rm f}^2\over|q_{\rm f}B|}\right)\right]+4(\slashed k_1+\slashed k_2)L_{n-1}^1\left({2{\bf k}_\bot^2\over|q_{\rm f}B|}\right),
\eea
where ${\cal P}_\pm^{\rm f}=1\pm{\rm sgn}(q_{\rm f} B)i\gamma^1\gamma^2$ is the spin up/down projector and $L_{n}^\alpha(x)$	are the generalized Laguerre polynomials with $L_{n}(x)\equiv L_{n}^0(x)$ and $L_{-1}^\alpha(x)=0$.
Then, the polarization loop for $\bar{\rho}^+_1$ meson with vanishing three-momentum can be evaluated as the following:
\bea\label{Pirho}
\Pi_{\bar{\rho}^+_1\bar{\rho}^+_1}(B,p_4)
&=&-32N_c\sum_{n=0}^\infty\sum_{n'=0}^\infty\int{\di^4k\over (2\pi)^4}e^{-{{\bf k}_\bot^2\over|q_{\rm u}B|}-{{\bf k}_\bot^2\over|q_{\rm d}B|}}{(m^2+k_3^2+(k_4+p_4)k_4)L_n\left({2{\bf k}_\bot^2\over|q_{\rm u}B|}\right)L_{n'}\left({2{\bf k}_\bot^2\over|q_{\rm d}B|}\right)\over ((k_4+p_4)^2+{E_{\rm u}^B}^2)(k_4^2+{E_{\rm d}^B}^2)}\nonumber\\
&=&-4N_c\sum_{n=0}^\infty\sum_{n'=0}^\infty{eB\over\pi}\int{\di {\bf k}_3\over (2\pi)}\left[{(m^2+E_{\rm u}^BE_{\rm d}^B+k_3^2)G_{nn'}\over p_4^2+(E_{\rm u}^B+E_{\rm d}^B)^2}\left({1\over E_{\rm u}^B}+{1\over E_{\rm d}^B}\right)\right],
\eea
\end{widetext}
where the quark dispersions in magnetic field are $E_{\rm u}^B\equiv \sqrt{k_3^2+m^2+2n|q_{\rm u} B|}$ and $E_{\rm d}^B\equiv \sqrt{k_3^2+m^2+2n'|q_{\rm d} B|}$, and the dimensionless $G$ function is defined as
\bea
{{G}}_{nn'}&\equiv&\int_0^\infty\di x~e^{-\left({1\over{|\tilde{q}_{\rm u}|}}+{1\over{|\tilde{q}_{\rm d}|}}\right)x}L_n\left({2x\over{|\tilde{q}_{\rm u}|}}\right)L_{n'}\left({2x\over{|\tilde{q}_{\rm d}|}}\right)\nonumber\\
&=&{1\over4}\sum_{k=0}^n\sum_{k'=0}^{n'}\left(\begin{array}{c}
		n\\n-k
\end{array}\right)\left(\begin{array}{c}
		n'\\n'-k'
\end{array}\right)\left(\begin{array}{c}
		k+k'\\k
\end{array}\right)\nonumber\\
&&(-2{|\tilde{q}_{\rm d}|})^{k+1}(-2{|\tilde{q}_{\rm u}|})^{k'+1}
	\eea
with the reduced charges $\tilde{q}_{\rm f}=q_{\rm f}/e$. As ${{G}}_{nn'}$ is magnetic field and energy-momentum independent, the matrix can be evaluated to very large $n$ and $n'$ numerically by utilizing {\it Mathematica} and then reserved as a special function for further manipulations. 

\begin{figure}[!htb]
	\begin{center}
		\includegraphics[width=8cm]{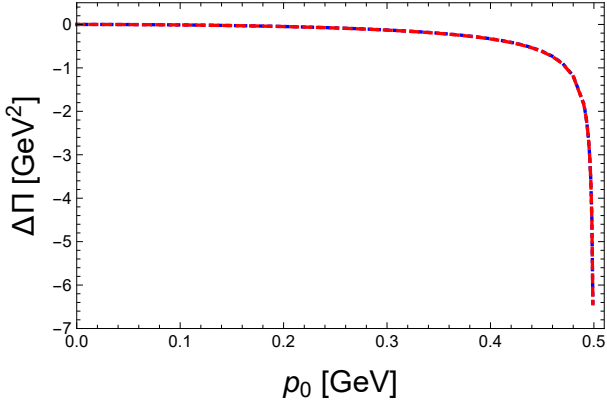}
		\caption{(color online) The comparison of $\Delta\Pi$ between proper-time (blue dotted line) and Landau-level (red dashed line) presentations for the chosen quark mass $m=0.25~{\rm GeV}$ and magnetic fields: $B_1=1~{\rm GeV}$ and $B_2=2~{\rm GeV}$. The result is very well convergent by increasing $n_{max}=n'_{max}$ from $100$ to $200$ for the Landau-level presentation.}\label{LLPT}
	\end{center}
\end{figure}

The bare polarization function Eq.\eqref{Pirho} is ultraviolet divergent and needs further regularization. We are not going to introduce any artificial cutoff at this stage for the purpose of demonstrating the equality between proper-time and Landau-level presentations, rather, the following formally convergent term is evaluated:
\bea
\!\!\!\!\!\!\!\!\Delta\Pi\equiv[\Pi_{\bar{\rho}^+_1\bar{\rho}^+_1}(B_2,i\,p_0)\!-\!\Pi_{\bar{\rho}^+_1\bar{\rho}^+_1}(B_2,0)]\!-\!(B_2\!\rightarrow\!B_1).
\eea
The comparison between proper-time (see Eq.\eqref{RPFLL1}) and Landau-level presentations is illuminated in Fig.~\ref{LLPT}, where they are found to be precisely consistent with each other up to the instable point $p_0=2m$. The equality should continue in the instable region $p_0>2m$, but $\Delta\Pi$ is not suitable for such exploration. The reason is that it diverges in the proper-time presentation and the variable transformation mentioned in Sec.~\ref{FLL} cannot be performed consistently to help as both $p_0=0$ and $p_0\neq0$ are involved now.

\section{Invalidity of NJL model in exploring the magnetic field effect to physical $\rho$ meson}\label{invalidity}
\begin{figure}[!htb]
	\begin{center}
		\includegraphics[width=8cm]{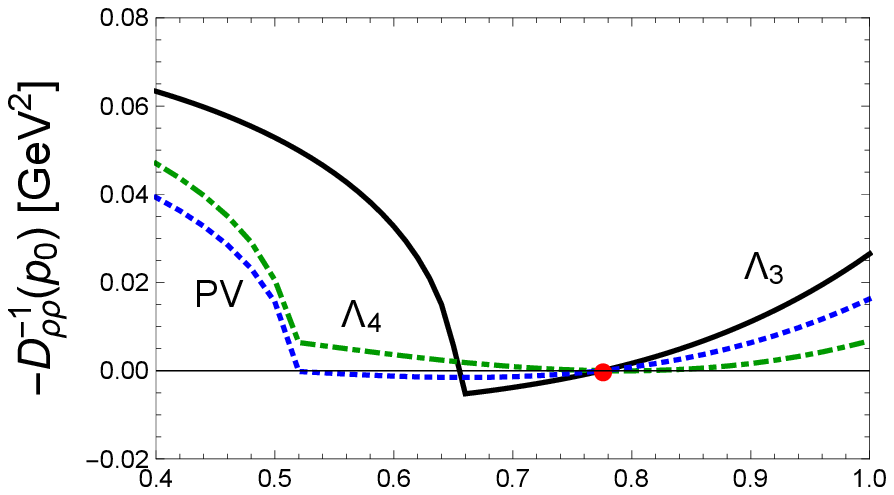}
		\includegraphics[width=8cm]{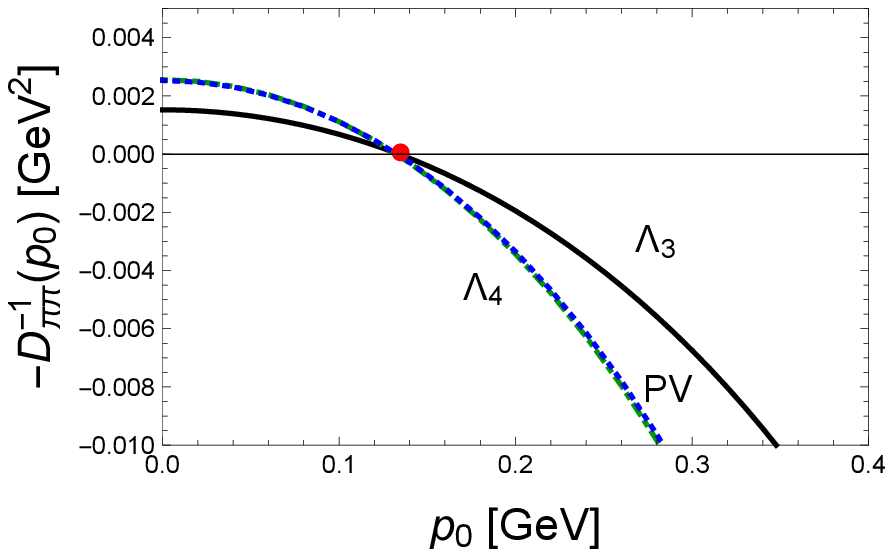}
		\caption{(color online) The effective inverse propagators of $\rho$ (upper panel) and $\pi^0$ (lower panel) mesons in vacuum with respect to different regularization schemes: three-momentum cutoff (black solid line), four-momentum cutoff (green dot-dashed line) and Pauli-Villars (blue dotted line). The red bullets are the physical $\rho$ and $\pi^0$ meson masses. In the upper panel, non-analytic features are developed at twice the corresponding dynamical quark masses: $0.313~{\rm GeV}$ for $\Lambda_3$ and $\sim 0.256~{\rm GeV}$ for $\Lambda_4$ and PV regularizations~\cite{Klevansky:1992qe}.}\label{prp_rhopi}
	\end{center}
\end{figure}

To explore $\rho$ meson property in certain circumstances, adequate regularization schemes should be chosen in NJL model first of all.
Here, we compare three regularization schemes with the parameters listed in Ref.~\cite{Klevansky:1992qe}: three-momentum cutoff ($\Lambda_3$), four-momentum cutoff ($\Lambda_4$) and Pauli-Villars (PV). In order to show the pion spectrum more explicitly, a non-vanishing current quark mass, $m_0=5~{\rm MeV}$, is adopted alternatively. Firstly, we study the spatial component $\rho_i$ of the vector rho meson in the absence of magnetic field and illuminate the results in the upper panel of Fig.~\ref{prp_rhopi} for the inverse propagators. 
As can be seen, there are two zero-points for all the regularization schemes, of which the other one is lighter than the physical mass in the $\Lambda_3$ and PV schemes but slightly heavier in the $\Lambda_4$ scheme. Recalling the basic form of vector boson propagator in quantum field theory (QFT)~\cite{Peskin}:
\bea
D_{VV}^{\mu\nu}(p)=-{g^{\mu\nu}-\hat{p}^\mu\hat{p}^\nu\over p^2-m^2},
\eea
we expect $-{D}^{-1}_{\bar{\rho}^+_1\bar{\rho}^+_1}>0~(<0)$ for $p_0<2m~(>2m)$. So, the signs of the $\rho$ meson propagators are wrong around the physical zero-point in the $\Lambda_3$ and PV schemes, and only $\Lambda_4$ scheme is suitable to describe the $\rho$ meson spectrum. 
For comparison, we show the inverse propagators of $\pi$ meson in the lower panel of Fig.~\ref{prp_rhopi}, where the curves are very close to each other for $\Lambda_4$ and PV schemes. All the inverse propagators share the same sign around their zero-point, which are consistent with the form of scalar boson propagator in QFT~\cite{Peskin}:
\bea
D_{SS}(p)={1\over p^2-m^2}.
\eea

\begin{figure}[!htb]
	\begin{center}
		\includegraphics[width=8cm]{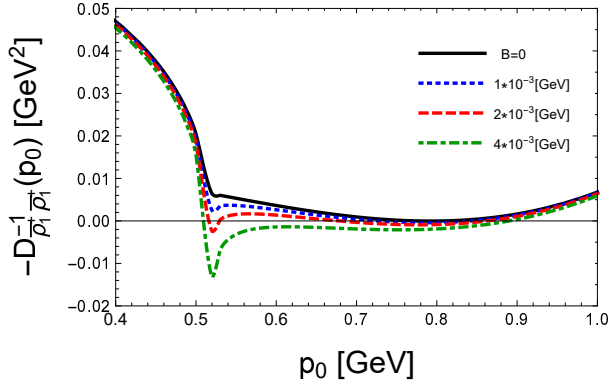}
		\caption{(color online) The effective inverse propagator of $\rho^+_1$ meson with respect to different values of magnetic field: $B=0$ (black solid line), $10^{-3}~{\rm GeV}^2$ (blue dotted line), $2*10^{-3}~{\rm GeV}^2$ (red dashed line) and $4*10^{-3}~{\rm GeV}^2$ (green dot-dashed line). There are dips around the two-quark instable point $p_0\sim2m$.}\label{prho_B}
	\end{center}
\end{figure}

Secondly, we study the magnetic effect to $\bar{\rho}^+_1$ meson by choosing the most optimistic $\Lambda_4$ scheme in  Eq.\eqref{RPFLLr}. The regularized terms can be given with the help of Feynman parameter as~\cite{Klimt:1989pm}:
\bea
&&-8N_c\int^{\Lambda_4}{\di^4k\over (2\pi)^4}{m^2\!+\!k_4(k_4+p_4)\!+\!k_3^2\over (k_4^2\!+\!E_{\bf k}^2)[(k_4+p_4)^2\!+\!E_{\bf k}^2]}\nonumber\\
&=&{N_c\over 6\pi^2}\left[-2\left(\Lambda^2-m^2\log\left(1+{\Lambda^2\over m^2}\right)\right)+(-p_4^2+2m^2)\right.\nonumber\\
&&\ \ \ \ \ \ \left.\int_0^1\di x\left({\Lambda^2F(x,\Lambda)}+\log\left(1-{\Lambda^2F(x,\Lambda)}\right)\right)\right],
\eea
\bea
&&-8N_c\int^{\Lambda_4}{\di^4k\over (2\pi)^4}{[m^2\!+\!k_4(k_4+p_4)\!+\!k_3^2]~eB\over (k_4^2\!+\!E_{\bf k}^2)^2[(k_4+p_4)^2\!+\!E_{\bf k}^2]}\nonumber\\
&=&-{N_ceB\over 4\pi^2}\!\left[\log\left(\!1\!+\!{\Lambda^2\over m^2}\!\right)\!+\!{p_4^2\over2}\!\!\int_0^1\!\!\di x\left(F(x,\Lambda)\!-\!F(x,0)\right)\right]\nonumber\\
\eea
with the auxiliary function $F(x,y)=[y^2-p_4^2(x^2-x)+m^2]^{-1}$. The numerical results are illuminated in Fig.\ref{prho_B}.
Contrary to the point particle results or LQCD simulations, so strong dips are developed around $p_0\sim2m$ in the spectra that $\bar{\rho}^+_1$ meson mass changes very quickly and discontinuously with $B$. This must be an artifact due to the absence of confinement in NJL model, because $\rho$ meson is not allowed to decay into quark-antiquark pair in the vacuum in real QCD. Even the formal extension to effectively include confinement through the Polyakov loop potential can not help the situation because it only plays an effective role at finite temperature~\cite{Ferreira:2013tba}. Thus, the conclusion is that NJL (or PNJL) model is not suitable to study the magnetic effect to heavy mesonic resonances with masses $>2m$, such as vectors $\rho$ and $\omega$, and pseudoscalar $\eta'$.


\begin{thebibliography}{99}
\bibitem{Kharzeev:2010gd} 
D.~E.~Kharzeev and H.~U.~Yee,
Phys.\ Rev.\ D {\bf 83}, 085007 (2011).

\bibitem{Son:2004tq} 
D.~T.~Son and A.~R.~Zhitnitsky,
Phys.\ Rev.\ D {\bf 70}, 074018 (2004).

\bibitem{Metlitski:2005pr} 
M.~A.~Metlitski and A.~R.~Zhitnitsky,
Phys.\ Rev.\ D {\bf 72}, 045011 (2005).

\bibitem{Huang:2013iia} 
X.~G.~Huang and J.~Liao,
Phys.\ Rev.\ Lett.\  {\bf 110}, no. 23, 232302 (2013).

\bibitem{Hattori:2016njk}
K.~Hattori and Y.~Yin,
Phys.\ Rev.\ Lett.\  {\bf 117}, no. 15, 152002 (2016).

\bibitem{Bali:2011qj}
G.~S.~Bali, F.~Bruckmann, G.~Endrodi, Z.~Fodor, S.~D.~Katz, S.~Krieg, A.~Schafer and K.~K.~Szabo,
JHEP {\bf 1202}, 044 (2012).



\bibitem{Bali:2012zg}
G.~S.~Bali, F.~Bruckmann, G.~Endrodi, Z.~Fodor, S.~D.~Katz and A.~Schafer,
Phys.\ Rev.\ D {\bf 86}, 071502(R) (2012).



\bibitem{Bruckmann:2013oba}
F.~Bruckmann, G.~Endrodi and T.~G.~Kovacs,
JHEP {\bf 1304}, 112 (2013).



\bibitem{Fukushima:2012kc}
K.~Fukushima and Y.~Hidaka,
Phys.\ Rev.\ Lett.\  {\bf 110}, no. 3, 031601 (2013).



\bibitem{Kojo:2012js}
T.~Kojo and N.~Su,
Phys.\ Lett.\ B {\bf 720}, 192 (2013).



\bibitem{Chao:2013qpa}
J.~Chao, P.~Chu and M.~Huang,
Phys.\ Rev.\ D {\bf 88}, 054009 (2013).

\bibitem{Cao:2014uva}
G.~Cao, L.~He and P.~Zhuang,
Phys.\ Rev.\ D {\bf 90}, no. 5, 056005 (2014).

\bibitem{Ferrer:2014qka}
E.~J.~Ferrer, V.~de la Incera and X.~J.~Wen,
Phys.\ Rev.\ D {\bf 91}, no. 5, 054006 (2015).

\bibitem{Mao:2016fha} 
S.~Mao,
Phys.\ Lett.\ B {\bf 758}, 195 (2016).

\bibitem{Chernodub:2010qx} 
M.~N.~Chernodub,
Phys.\ Rev.\ D {\bf 82}, 085011 (2010).

\bibitem{Chernodub:2011mc} 
M.~N.~Chernodub,
Phys.\ Rev.\ Lett.\  {\bf 106}, 142003 (2011).

\bibitem{Braguta:2011hq} 
V.~V.~Braguta, P.~V.~Buividovich, M.~N.~Chernodub, A.~Y.~Kotov and M.~I.~Polikarpov,
Phys.\ Lett.\ B {\bf 718}, 667 (2012).

\bibitem{Liu:2014uwa} 
H.~Liu, L.~Yu and M.~Huang,
Phys.\ Rev.\ D {\bf 91}, no. 1, 014017 (2015).

\bibitem{Hidaka:2012mz} 
Y.~Hidaka and A.~Yamamoto,
Phys.\ Rev.\ D {\bf 87}, no. 9, 094502 (2013).

\bibitem{Bali:2017ian} 
G.~S.~Bali, B.~B.~Brandt, G.~Endrodi and B.~Glassle,
Phys.\ Rev.\ D {\bf 97}, no. 3, 034505 (2018).

\bibitem{Skokov:2009qp}
V.~Skokov, A.~Y.~Illarionov and V.~Toneev,
Int.\ J.\ Mod.\ Phys.\ A {\bf 24}, 5925 (2009).



\bibitem{Deng:2012pc}
W.~T.~Deng and X.~G.~Huang,
Phys.\ Rev.\ C {\bf 85}, 044907 (2012).



\bibitem{Deng:2014uja}
W.~T.~Deng and X.~G.~Huang,
Phys.\ Lett.\ B {\bf 742}, 296 (2015).

\bibitem{Bloczynski:2012en}
J.~Bloczynski, X.~G.~Huang, X.~Zhang and J.~Liao,
Phys.\ Lett.\ B {\bf 718}, 1529 (2013).

\bibitem{Guo:2019joy} 
Y.~Guo, S.~Shi, S.~Feng and J.~Liao,
arXiv:1905.12613 [nucl-th].

\bibitem{Liao:2014ava}
J.~Liao,
Pramana {\bf 84}, no. 5, 901 (2015).



\bibitem{Kharzeev:2015znc}
D.~E.~Kharzeev, J.~Liao, S.~A.~Voloshin and G.~Wang,
Prog.\ Part.\ Nucl.\ Phys.\  {\bf 88}, 1 (2016).



\bibitem{Huang:2015oca}
X.~G.~Huang,
Rept.\ Prog.\ Phys.\  {\bf 79}, no. 7, 076302 (2016).

\bibitem{Adamczyk:2014mzf} 
L.~Adamczyk {\it et al.} [STAR Collaboration],
Phys.\ Rev.\ Lett.\  {\bf 113}, 052302 (2014).

\bibitem{Zhao:2017nfq} 
J.~Zhao, H.~Li and F.~Wang,
Eur.\ Phys.\ J.\ C {\bf 79}, no. 2, 168 (2019).

\bibitem{Magdy:2017yje} 
N.~Magdy, S.~Shi, J.~Liao, N.~Ajitanand and R.~A.~Lacey,
Phys.\ Rev.\ C {\bf 97}, no. 6, 061901(R) (2018).

\bibitem{Miransky:2015ava}
V.~A.~Miransky and I.~A.~Shovkovy,
Phys.\ Rept.\  {\bf 576}, 1 (2015).

\bibitem{Frolov:2010wn} 
I.~E.~Frolov, V.~C.~Zhukovsky and K.~G.~Klimenko,
Phys.\ Rev.\ D {\bf 82}, 076002 (2010).

\bibitem{Cao:2016fby} 
G.~Cao and A.~Huang,
Phys.\ Rev.\ D {\bf 93}, no. 7, 076007 (2016).

\bibitem{Cao:2015cka} 
G.~Cao and X.~G.~Huang,
Phys.\ Lett.\ B {\bf 757}, 1 (2016).

\bibitem{Wang:2017pje} 
L.~Wang and G.~Cao,
Phys.\ Rev.\ D {\bf 97}, no. 3, 034014 (2018).

\bibitem{Wang:2018gmj} 
L.~Wang, G.~Cao, X.~G.~Huang and P.~Zhuang,
Phys.\ Lett.\ B {\bf 780}, 273 (2018).

\bibitem{Wang:2017vtn} 
Z.~Wang and P.~Zhuang,
Phys.\ Rev.\ D {\bf 97}, no. 3, 034026 (2018).

\bibitem{Mao:2018dqe} 
S.~Mao,
Phys.\ Rev.\ D {\bf 99}, no. 5, 056005 (2019).

\bibitem{Liu:2018zag} 
H.~Liu, X.~Wang, L.~Yu and M.~Huang,
Phys.\ Rev.\ D {\bf 97}, no. 7, 076008 (2018).

\bibitem{Avancini:2016fgq} 
S.~S.~Avancini, R.~L.~S.~Farias, M.~Benghi Pinto, W.~R.~Tavares and V.~S.~Timóteo,
Phys.\ Lett.\ B {\bf 767}, 247 (2017).

\bibitem{Coppola:2018vkw} 
M.~Coppola, D.~Gómez Dumm and N.~N.~Scoccola,
Phys.\ Lett.\ B {\bf 782}, 155 (2018).

\bibitem{Shushpanov:1997sf} 
I.~A.~Shushpanov and A.~V.~Smilga,
Phys.\ Lett.\ B {\bf 402}, 351 (1997).

\bibitem{Agasian:2001ym} 
N.~O.~Agasian and I.~A.~Shushpanov,
JHEP {\bf 0110}, 006 (2001).

\bibitem{Hengtong2019} 
Hengtong Ding's talk in the workshop on "Chirality, Vorticity and Magnetic Field in Heavy Ion Collisions", 
Beijing (2019).

\bibitem{Ambjorn:1988tm} 
J.~Ambjorn and P.~Olesen,
Nucl.\ Phys.\ B {\bf 315}, 606 (1989).

\bibitem{Ambjorn:1988gb} 
J.~Ambjorn and P.~Olesen,
Phys.\ Lett.\ B {\bf 218}, 67 (1989)
Erratum: [Phys.\ Lett.\ B {\bf 220}, 659 (1989)].

\bibitem{Chernodub:2012fi} 
M.~N.~Chernodub, J.~V.~Doorsselaere and H.~Verschelde,
Phys.\ Rev.\ D {\bf 88}, 065006 (2013).

\bibitem{Cao:2015xja} 
G.~Cao and P.~Zhuang,
Phys.\ Rev.\ D {\bf 92}, no. 10, 105030 (2015).

\bibitem{Klevansky:1992qe}
S.~P.~Klevansky,
Rev.\ Mod.\ Phys.\  {\bf 64}, 649 (1992).

\bibitem{Klimt:1989pm} 
S.~Klimt, M.~F.~M.~Lutz, U.~Vogl and W.~Weise,
Nucl.\ Phys.\ A {\bf 516}, 429 (1990).


\bibitem{Schwinger:1951nm}
J.~S.~Schwinger,
Phys.\ Rev.\  {\bf 82}, 664 (1951).
	
\bibitem{Ferreira:2013tba} 
M.~Ferreira, P.~Costa, D.~P.~Menezes, C.~Providencia and N.~N.~Scoccola,
Phys.\ Rev.\ D {\bf 89}, no. 1, 016002 (2014)
Addendum: [Phys.\ Rev.\ D {\bf 89}, no. 1, 019902 (2014)].


\bibitem{Brauner:2016lkh} 
T.~Brauner and X.~G.~Huang,
Phys.\ Rev.\ D {\bf 94}, no. 9, 094003 (2016).

\bibitem{Zhuang:1994dw}
P.~Zhuang, J.~Hufner and S.~P.~Klevansky,
Nucl.\ Phys.\ A {\bf 576}, 525 (1994).

\bibitem{tHooft:1976snw}
G.~'t Hooft,
Phys.\ Rev.\ D {\bf 14}, 3432 (1976)
Erratum: [Phys.\ Rev.\ D {\bf 18}, 2199 (1978)].

\bibitem{Rehberg:1995kh}
P.~Rehberg, S.~P.~Klevansky and J.~Hufner,
Phys.\ Rev.\ C {\bf 53}, 410 (1996).

\bibitem{Liu:2017spl} 
Y.~Liu and I.~Zahed,
Phys.\ Rev.\ Lett.\  {\bf 120}, no. 3, 032001 (2018).

\bibitem{CaoandChen}
G. Cao and L. He, arXiv:1910.02728; H.-L. Chen, X.-G. Huang, and K. Mameda, arXiv:1910.02700.

\bibitem{Peskin}
M.E. Peskin and D.V. Schroeder, {\it An Introduction to Quantum Field Theory} (Addison-Wesley, Reading, MA, 1995). 

\end{thebibliography}
\end{document}